\documentclass{aa}  

\usepackage{graphicx}
\usepackage{natbib}
\usepackage{hyperref}
\bibpunct{(}{)}{;}{a}{}{,} % to follow the A&A style
\usepackage[varg]{txfonts}
\def\kms{$\rm km\, s^{-1}$}

\def\fe2{[Fe\,{\sc ii}]}

\def\o3{O\,[{\sc iii}]}
\def\h2{H$_{2}$}

\def\cm3{$\rm cm^{-3}$}

\def\s3{[S{\sc iii}]}

\def\kms    {\ifmmode{{\rm \ts km\ts s}^{-1}}\else{\ts km\ts s$^{-1}$}\fi}
\def\msol   {\ifmmode{{\rm M}_{\odot}}\else{M$_{\odot}$}\fi}
\def\lsol   {\ifmmode{{\rm L}_{\odot}}\else{L$_{\odot}$}\fi}
\def\zsol   {\ifmmode{{\rm Z}_{\odot}}\else{Z$_{\odot}$}\fi}

\begin{document}

   \title{ALMA captures feeding and feedback from the active galactic nucleus in NGC\,613}
 %\subtitle{}
  \author{A. Audibert \inst{1}  \and F. Combes\inst{1,}\inst{2} \and S. Garc{\'i}a-Burillo\inst{3} \and L. Hunt \inst{4} \and A. Eckart \inst{5}  \and S. Aalto \inst{6} \and V. Casasola \inst{7} \and  F. Boone \inst{8} \and M. Krips \inst{9} \and S. Viti \inst{10}  \and S. Muller \inst{6} \and K. Dasyra \inst{11}  \and P. van der Werf \inst{12} \and S. Mart{\'i}n\inst{13,} \inst{14}}

   \institute{Observatoire de Paris, LERMA, CNRS, PSL Univ., Sorbonne University, UPMC, Paris, France\\
              \email{anelise.audibert@obspm.fr}
         \and
			Coll{\`e}ge de France, 11 Pl. Marcelin Berthelot, 75231, Paris  
         \and
             Observatorio Astron{\'o}mico Nacional (OAN-IGN)-Observatorio de Madrid, Alfonso XII, 3, 28014 Madrid, Spain
		\and
			 INAF - Osservatorio Astrofisico di Arcetri, Largo E. Fermi, 5, 50125, Firenze, Italy \and I. Physikalisches Institut, Universit{\"a}t zu K{\"o}ln, Z{\"u}lpicher Str. 77, 50937, K{\"o}ln, Germany
\and Department of Space, Earth and Environment, Chalmers University of Technology, Onsala Space Observatory, SE-43992 Onsala, Sweden
\and INAF - Isitituto di Radioastronomia, via Piero Gobetti 101, 40129, Bologna, Italy
\and CNRS, IRAP, 9 Av. colonel Roche, BP 44346, 31028, Toulouse Cedex 4, France
\and IRAM, 300 rue de la Piscine, Domaine Universitaire, F-38406 Saint Martin d’H{\`e}res, France
\and Dep. of Physics and Astronomy, UCL, Gower Place, London WC1E 6BT, UK
\and Dep. of Astrophysics, Astronomy \& Mechanics, Faculty of Physics, National and Kapodistrian University of Athens, Panepistimiopolis Zografou, 15784, Greece, and National Observatory of Athens, Institute for Astronomy, Astrophysics, Space
Applications and Remote Sensing, Penteli, 15236, Athens, Greece
\and Leiden Observatory, Leiden Univ., PO Box 9513, 2300 RA Leiden, Netherlands
\and European Southern Observatory, Alonso de C{\'o}rdova, 3107, Vitacura, Santiago 763-0355, Chile
\and Joint ALMA Observatory, Alonso de C{\'o}rdova, 3107, Vitacura, Santiago 763-0355, Chile  \\ }        

   \date{Received xx xx, 2019; accepted xxxx}

% \abstract{}{}{}{}{} 
% 5 {} token are mandatory
 
\abstract{We report ALMA observations of CO(3-2) emission in the Seyfert/nuclear starburst galaxy NGC\,613, at a spatial resolution of 17\,pc, as part of our NUclei of GAlaxies (NUGA) sample. Our aim is to investigate the morphology and dynamics of the gas inside the central kpc, and to probe nuclear fueling and feedback phenomena. The morphology of CO(3-2) line emission reveals a 2-arm trailing nuclear spiral at $\rm r\lesssim$100\,pc and a circumnuclear ring at $\sim$350\,pc radius, that is coincident with the star-forming ring seen in the optical images. Also, we find evidence of a filamentary structure connecting the ring and the nuclear spiral. The ring reveals two breaks into two winding spiral arms corresponding to the dust lanes in the optical images. The molecular gas in the galaxy disk is in a remarkably regular rotation, however, the kinematics in the nuclear region is very skewed. The nuclear spectrum of CO and dense gas tracers $\rm HCN$(4-3), $\rm HCO^+$(4-3), and $\rm CS$(7-6) show broad wings up to $\pm$300\,km/s, associated with a molecular outflow emanating from the nucleus ($r\sim$25\,pc). We derive a molecular outflow mass $M_{out}$=2$\times$10$^6$M$_\odot$ and a mass outflow rate of  $\dot{M}_{out}=$27\,$\rm M_\odot yr^{-1}$. The molecular outflow energetics exceed the values predicted by AGN feedback models: the kinetic power of the outflow corresponds to $P_{K,out}=$20\%$L_{AGN}$ and the momentum rate is $\dot{M}_{out}v\sim400L_{AGN}/c$. The outflow is mainly boosted by the AGN through entrainment by the radio jet, but given the weak nuclear activity of NGC\,613, we might be witnessing a \textit{fossil outflow}, resulted from a strong past AGN that now has already faded.  Furthermore, the nuclear trailing spiral observed in CO emission is inside the inner Lindblad resonance (ILR) ring of the bar. We compute the gravitational torques exerted in the gas to estimate the efficiency of the angular momentum exchange. The gravity torques are negative from 25 to 100\,pc and the gas loses its angular momentum in a rotation period, providing evidence of a highly efficient inflow towards the center. This phenomenon shows that the massive central black hole has a significant dynamical influence on the gas, triggering the inflowing of molecular gas to feed the black hole. 
}


   \maketitle
%
%________________________________________________________________

\section{Introduction}

%__________________________________________________________________

The energy of active galactic nuclei (AGN) is well interpreted as due to gas accretion onto the supermassive black hole (SMBH) \citep{antonucci93}. Gas inflows into the center of galaxies can fuel the SMBH and the energy input  by the AGN can trigger subsequent feedback. The feedback can in turn regulate the SMBH growth and suppress star formation \citep[e.g.][]{croton06,sij07}. Feeding and feedback are key processes to understand the co-evolution of black holes (BH) and their host galaxies, which is now well established by the tight M-$\sigma$ relation \citep[e.g.][]{mag98,gul09,mcma13}. It is important to study the efficiency of angular momentum transport in galaxy disks in order to understand how the star formation and nuclear activity are fueled and what are the timescales involved, since both feeding processes rely on a common cold gas supply, but in different periods of time \citep[$\sim$10$^5$\,yr for BH growth and $\sim$10$^{7- 9}$\,yr for star formation,][]{santi16}. These time scales are related to the mechanisms that drive the gas from galactic scales ($\sim$10\,kpc) to nuclear scales (a few pc), through removal of angular momentum; large non-axisymmetric perturbations, such as bars or spirals, can do the job, but there is not yet identified a unique physical process associated with inward transport of gas in galaxy disks.

On large scales, cosmological simulations show that mergers and galaxy interactions are able to produce strong non-axisymmetries \citep[e.g.][]{hop06,dima08}. At kpc scales, bar instabilities, either internally driven by secular evolution or triggered by companions, can first feed a central starburst and then fuel the BH \citep{santi05}. On the other hand, gas inflow is impeded by the inner Lindblad resonance (ILR), where the gas is trapped in a nuclear ring \citep[see][]{piner95,regan04}. At few hundreds of pc scales, the ``bars within bars” scenario \citep[e.g.][]{shlo89}, together with m=1 instabilities and nuclear warps \citep{eva00} take over as a dynamical mechanism \citep[see e.g.,][]{hunt08}. Observations of nearby low luminosity AGN (LLAGN) with the NUGA program have revealed smoking-gun evidence of AGN fueling in one third of the galaxies \citep{santi12}. This result suggests that galaxies alternate periods of fueling and starvation, and might be found in a feeding phase at 300\,pc scales only one third of the time. 

As we approach the center of galaxies, other mechanisms could contribute to the fueling: viscous torques, from dense gas in regions of large shear, or dynamical friction can drive massive clouds to the nucleus \citep[e.g.][]{combes03,jogee06}. Simulations suggest that fueling involves a series of dynamical instabilities (m=2, m=1) at $\sim$10\,pc scales, and also predict the formation of a thick gas disk similar to the putative torus invoked to explain obscured AGN \citep{hop10, hop12}. These fueling episodes are eventually quenched by either nuclear star formation winds or AGN feedback. 

The recent discovery of many massive (a few 10$\rm ^7\,M_\odot$) molecular outflows in nearby AGN \citep[eg.,][]{fischer10,fer10,fer17,ala11, sturm11,vei13,cicone14, saka14, santi1068, kal12,kal14} has promoted the idea that winds may be major actors in sweeping gas out of galaxies. It has been already established that the mass outflow rates increase with the AGN luminosity, supporting the idea of a luminous AGN pushing away the surrounding gas through a fast wind. Observational works \citep{cicone14,fio17,flu19} have shown that the molecular outflow properties are correlated with the AGN luminosity, where the outflow kinetic power corresponds to about 5\%$\rm L_{AGN}$ and the momentum rate is $\sim$20$\rm L_{AGN}$/c, in agreement with theoretical models of AGN feedback \citep{fau12, zub12, zub14}. Outflows have been traced for a long time in ionized or atomic gas \citep{rupke05, rogemar11}, making it now possible to compare the different gas phases. \citet{carni15} found that ionised gas only traces a small fraction of the total gas mass, suggesting that the molecular phase dominates the outflow mass. This trend is also found by \citet{fio17}, but the ratio between molecular to ionised mass outflow rates is reduced at the highest AGN bolometric luminosities.
 
Probing AGN feeding and feedback phenomena through the kinematic and morphology of the gas inside the central kpc has only recently been possible due to the unprecedented Atacama Large Millimeter/submillimeter Array (ALMA) spatial resolution and sensitivity. Evidence of AGN feeding was found in NGC\,1566, where a molecular trailing spiral structure from 50 to 300\,pc was detected with ALMA Cycle 0 observations, and according to its negative gravity torques, is contributing to fuel the central BH \citep{combes1566}. Also with Cycle 0 observations, a molecular outflow is also seen a the LLAGN in the Seyfert\,2 NGC\,1433. It is the least massive molecular outflow ($\rm \sim4\times10^6\,M_\odot$) ever detected around galaxy nuclei \citep{combes1433}. A fast and collimated outflow has been detected in HCN(1-0) and CO(1-0) emission in the nucleus of Arp\,220, extending up to 120\,pc and reaching velocities up to $\pm$840 km/s \citep{barcos18}. 

In the prototypical Seyfert 2 NGC\,1068, a clear molecular ouflow has also been detected, entrained by the AGN radio jets \citep{krips11, santi1068}. NGC\,1068 is also the first case to resolve the molecular torus, using the continuum and the CO(6-5), HCN(3-2) and HCO$^+$(3- 2) emission lines observed with ALMA \citep{galli16,santi16tor,ima16}. The dynamics of the molecular gas in the NGC 1068 torus revealed strong non-circular motions and enhanced turbulence superposed on a slow rotation pattern of the disk. The AGN is clearly off-center with respect to the torus, implying an m=1 perturbation \citep{santi16tor}. Recently, we have reported observations of molecular tori around massive BHs in a sample of 7 nearby LLAGN (Seyfert/LINER), at the unprecedented spatial resolution of 3-10\,pc \citep[][, hereafter Paper I]{combes18}. The ALMA observations bring a wealth of new information on the decoupled molecular tori, which are found to have radii ranging from 6 to 27 pc, unaligned with the orientation of the host galaxy and frequently slightly off-centered to the AGN position. The kinematics of the gas inside the sphere of influence (SoI) of the central BH also allowed us to estimate the BH masses ($\rm M_{BH}\sim10^{7-8}\,M_\odot$).
 
In this paper, we present the combined ALMA cycle 3 and 4 observations in the CO(3-2) line of the Seyfert galaxy NGC\,613, with a spatial resolution of 17\,pc. These observations were part of the sample presented in Paper I, but the data analyzed here have more sensitivity in order to detect faint broad wings usually associated to outflows. This object presents a nuclear trailing spiral and we discovered a molecular outflow in its nuclear region. Therefore, NGC\,613 is a special case to study the complexity of fuelling and feedback mechanisms in AGN, and the detailed analysis of the gas flow cycle in AGN. In the next subsection, all relevant characteristics of NGC\,613 are described. Observations are detailed in Section~\ref{obs} and results in Section~\ref{results}. The properties of the nuclear molecular outflow discovered in the very center of NGC\,613 are discussed in Section~\ref{outflow}. The interpretation of inflowing gas in term of gravitational torques is discussed in Section~\ref{torque}, and conclusions are drawn in Section \ref{con}.

\begin{figure}
\resizebox{\hsize}{!}{\includegraphics{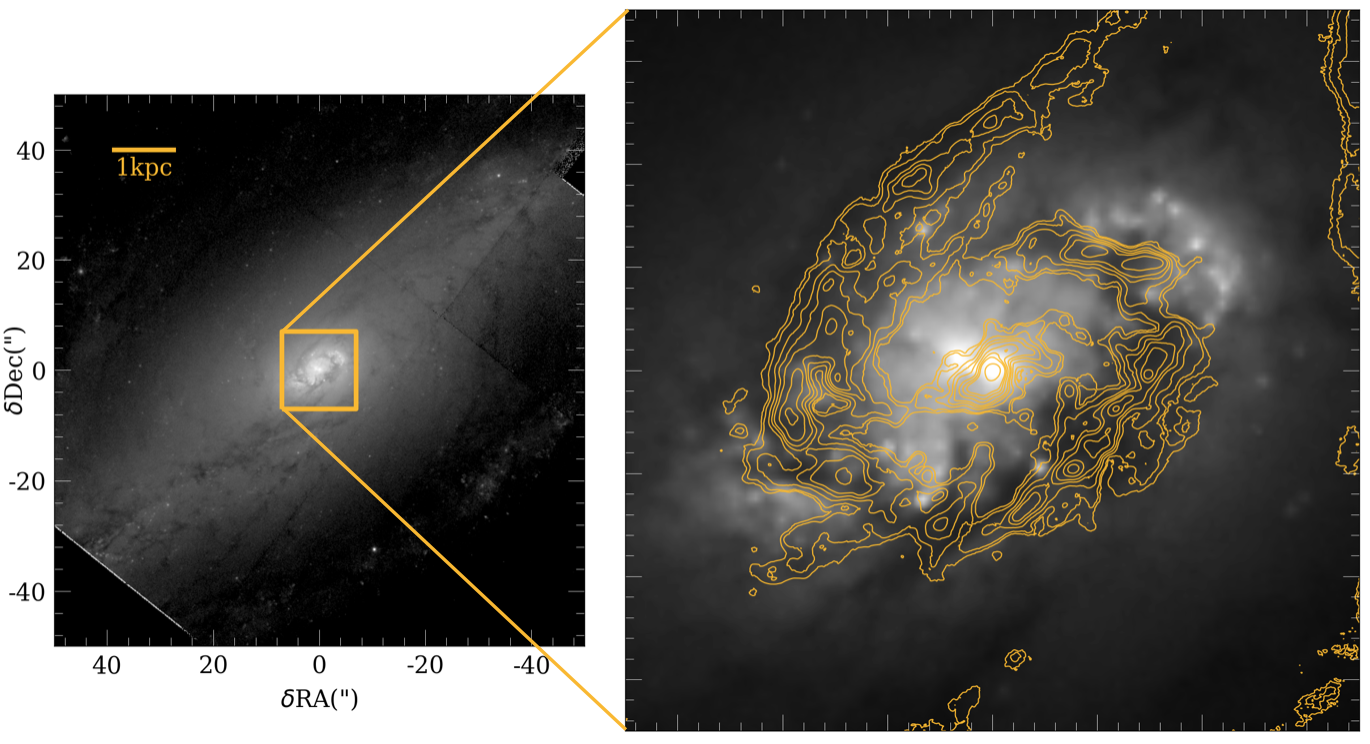}}
\caption{WFC/F814W HST image in the left panel, showing the nuclear ring and the  sets of spiral arm structures at large scales. The scale is indicated in the upper-left corner. \textit{Right:}  a 14\arcsec square  zoom showing the CO(3-2) contours overlaid on the same image. The HST image has been aligned to match the ALMA astrometry.}
\label{fig:hst}
\end{figure}

\begin{table}
\caption{Properties of NGC\,613}              % title of Table
\centering                                      
\begin{tabular}{l c c}         
\hline\hline   
\noalign{\smallskip}                    
 Parameter &Value & Reference \\  
\hline    
\noalign{\smallskip}                            
${\alpha_{J2000}}^a$ & 01h34m18.23s   & (1) \\
${\delta_{J2000}}^a$ &  -29d25m06.56s & (1) \\
V$\rm _{hel}$ & 1481$\pm$5\,km$\rm s^{-1}$ &  (1) \\
RC3 Type & SB(rs)bc & (1) \\
Nuclear Activity  & Sy/ Starburst & (1,2) \\
Inclination  & 41$^\circ$ & (3)  \\
Position Angle  & 120$^\circ$ & (3)  \\
Distance$^b$ & 17.2\,Mpc & (1) \\
SFR$^c$ & 5.3$\rm\,M_\odot yr^{-1}$ & (1) \\
%L$\rm _{B}$ &  & \\
$\rm M_{H\textsc{i}}$ & 4.7$\rm \times10^{9}\,M_\odot$ & (4) \\
$\rm M_*$ - Stellar Mass & 4.5$\rm \times10^{10}\,M_\odot$ & (5)  \\
%$\rm M_{dust}$(60 and 100$\mu$m) &  & \\
$\rm L_{IR}$ & 3$\times$10$\rm ^{10}L_\odot$ & (6) \\
${\alpha_{J2000}}^d$ & 01h34m18.19s  &  (7) \\
${\delta_{J2000}}^d$ & -29d25m06.59s & (7)  \\
\hline        
\label{tab:prop}
\end{tabular}
\tablefoot{References:  1: NASA/IPAC Extragalactic Database (NED); 2: \citet{veron86}; 3:  \citep{vau91}; 4: \citet{gado19}; 5: \citet{combes18}; 6: \citep{sturm02}; 7: this work. \\
\tablefoottext{$^a$}{($\alpha_{J2000}$, $\delta_{J2000}$) is the phase tracking center of our interferometric observations.} \\
\tablefoottext{$^b$}{Distance is the median values of z-independent distances from NED \citet{steer17} } \\
\tablefoottext{$^c$}{} SFR is derived from infrared luminosities (NED) \\
\tablefoottext{$^d$}{The RA-DEC positions are the new adopted center, derived from the central continuum peak in this work, with an uncertainty of $\sim$0.1" (see Sec.~\ref{cont_emission})} \\
}
\end{table}

\subsection{NGC\,613}

NGC\,613 is a nearby barred SB(rs)bc galaxy \citep{vau91} at a distance of 17.2\,Mpc (1\arcsec=83\,pc).  It has a large-scale bar of $\rm r_{bar}\sim$90\arcsec with a position angle of 127$\rm^\circ$ and a secondary nuclear bar with PA=122$\rm ^{\circ}$ \citep{jun97,seigar18}. Judging from NED, NGC\,613 is only moderately inclined, with inclination $\sim\,38^\circ$ (see also Section~\ref{kin}). Prominent dust lanes are visible along the large-scale bar and the presence of multiple spirals arms gives it a ``tentacular'' appearance. NGC\,613 has a typical inner Lindblad resonance (ILR) nuclear ring, of radius $\sim$3.5\arcsec (300 pc), just inside the two characteristic leading dust lanes of the bar (Fig~\ref{fig:hst}).

It hosts a LLAGN, that first was classified as a composite Seyfert/H\textsc{ii} object by \citet{veron86} based on its low-resolution optical spectrum, and later confirmed through MIR spectroscopy \citep{gou09} and X-ray observations using the \textit{ROSAT} and XMM-Newton \citep[][respectively]{liu05,casta13}. Water masers have been detected in the nucleus by \citet{kon06}.

NGC\,613 shows clear evidence of star formation, shock excitation and AGN activity \citep{davies17}. Radio continuum observations show evidence for a collimated jet from the AGN and a nuclear ring with a moderate inclination of i$\sim$55$^\circ$ \citep{hum87,hum92}. The presence of the outflow has already been suggested by the high velocity dispersion of the [Fe\textsc{ii}] line along the radio jet \citep{fal14}. The ring-like structure in Br$\gamma$ emission in NGC\,613, comprising  ``hot spots'' of current massive star formation \citep{boker08,fal14}, indicates an ongoing star-forming episode. We collect the main properties of NGC 613 in Table~\ref{tab:prop}.

\section{Observations}\label{obs}

We report the combined ALMA Cycle 3 and Cycle 4 observations of CO(3-2), CS(7-6), HCN(4-3) and HCO$\rm^{+}$(4-3) and continuum in band 7, at rest frame frequencies $\nu_{rest}$ of 345.8, 342.9, 354.5, 356.7 and 350\,GHz, respectively. The observations of the NUGA sample are described in Paper I; here we include additional details for NGC\,613.

In Cycle 3, NGC\,613 was observed (project ID: \#2015.1.00404.S, PI F. Combes) simultaneously in CO(3-2), HCO$\rm^{+}$(4-3), HCN(4-3) for both the compact (TC, baselines 15 to 630m) and the extended (TE, baselines 15 to 1400m) configurations. The largest recoverable angular scale corresponding to the shortest baseline is about 12\arcsec. The TC configuration was observed in April 2016 with 40 antennas and an integration time, including calibration and overheads, of 20 minutes, providing a synthesized beam of $\sim$0\farcs38. The TE configuration was observed in August 2016 with 41 antennas, total integration of 40 minutes and a synthesised beam of $\sim$0\farcs14. The correlator setup, designed to simultaneously observe three lines, provided a velocity range of 1600\,km/s for each line, but did not center the HCO$\rm^{+}$(4-3) and HCN(4-3) lines (200\,km/s on one side and 1400\,km/s on the other, which is adequate for a nearly face-on galaxy), and 1800\,MHz bandwidth in the continuum.

The Cycle 4 observation were carried out in November 2016 and July 2017 (project ID: \#2016.1.00296.S, PI F. Combes) at higher spatial resolution ($\sim$6.5\,pc) aiming at resolving the molecular torus. The tuning configuration of Band 7 was in the CO(3-2), HCO$\rm^+$(4-3) and continuum, to avoid a restricted velocity range in the expected broader spectral lines towards the nucleus. The correlator setup was selected to center the CO(3-2) and the HCO$\rm ^+$ lines in the 2\,GHz bandwidth. The compact configuration (TM2, baselines 19 to 500m) was observed with 44 antennas for an integration time of 14 minutes and a synthesised beam of 0\farcs31 and the extended (TM1, baselines 19 to 3100\,m) with 43 antennas for 1.2 hours and a synthesised beam of $\sim$0\farcs08. 

The observations phase center was that of the nucleus (Table~\ref{tab:prop}), with a single pointing covering a Field of View (FoV) of 18\arcsec. The galaxy was observed in dual polarization mode with 1.875 GHz total bandwidth per spectral window, and a channel spacing of 0.488 MHz corresponding to $\sim$0.8 km/s, after Hanning smoothing. The flux calibration was done with radio quasars close to the position in the sky of the target, which are regularly monitored at ALMA, and resulted in 10\% accuracy.

The data from Cycle 3 and 4 were calibrated and concatenated with the \textsc{CASA} software (version from 4.5.3 to 4.7.2, \citep{casa}, and the imaging and cleaning were performed with the \textsc{GILDAS} software \citep{gildas}. In Paper I, we have used only the most extended configurations (TM1+TE), but in this work we have combined all the configurations to improve the sensitivity. The analysis were made in \textsc{GILDAS} together with \textsc{python} packages \citep[radio-astro-tools, APLpy, PySpecKit][]{radiotools, aplpy,pyspec}. The \textsc{CLEAN}ing was performed using the Hogbom method and a natural weighting, in order to achieve the best sensitivity, resulting in a synthesised beam of 0\farcs21$\times$0\farcs19 for the concatenated data cube. The spectral line maps were obtained after subtraction of the continuum in the $uv$-plane using the tasks \textsc{uv\_continuum} and \textsc{uv\_subtract}. The data cubes were then smoothed to 10\,km/s (11.5\,MHz). The total integration time provided an rms of 87$\mu$Jy/beam in the continuum, and in the line channel maps 0.43\,mJy/beam per channel of 10\,km/s (corresponding to $\sim$1K, at the obtained spatial resolution). The final maps were corrected for primary beam attenuation. Very little CO(3-2) emission was detected outside the full-width half-power (FWHP) primary beam. 

Because of lack of very short baselines ($<$15m), extended emission was filtered out at scales larger than 12\arcsec~ in each channel map. Since the velocity gradients are high in galaxy nuclei, this does not affect significantly the line measurements: indeed, the size in each velocity channel is not expected to be extended. %However, the missing-flux problem might affect continuum maps. 

\begin{figure}
\resizebox{\hsize}{!}{\includegraphics{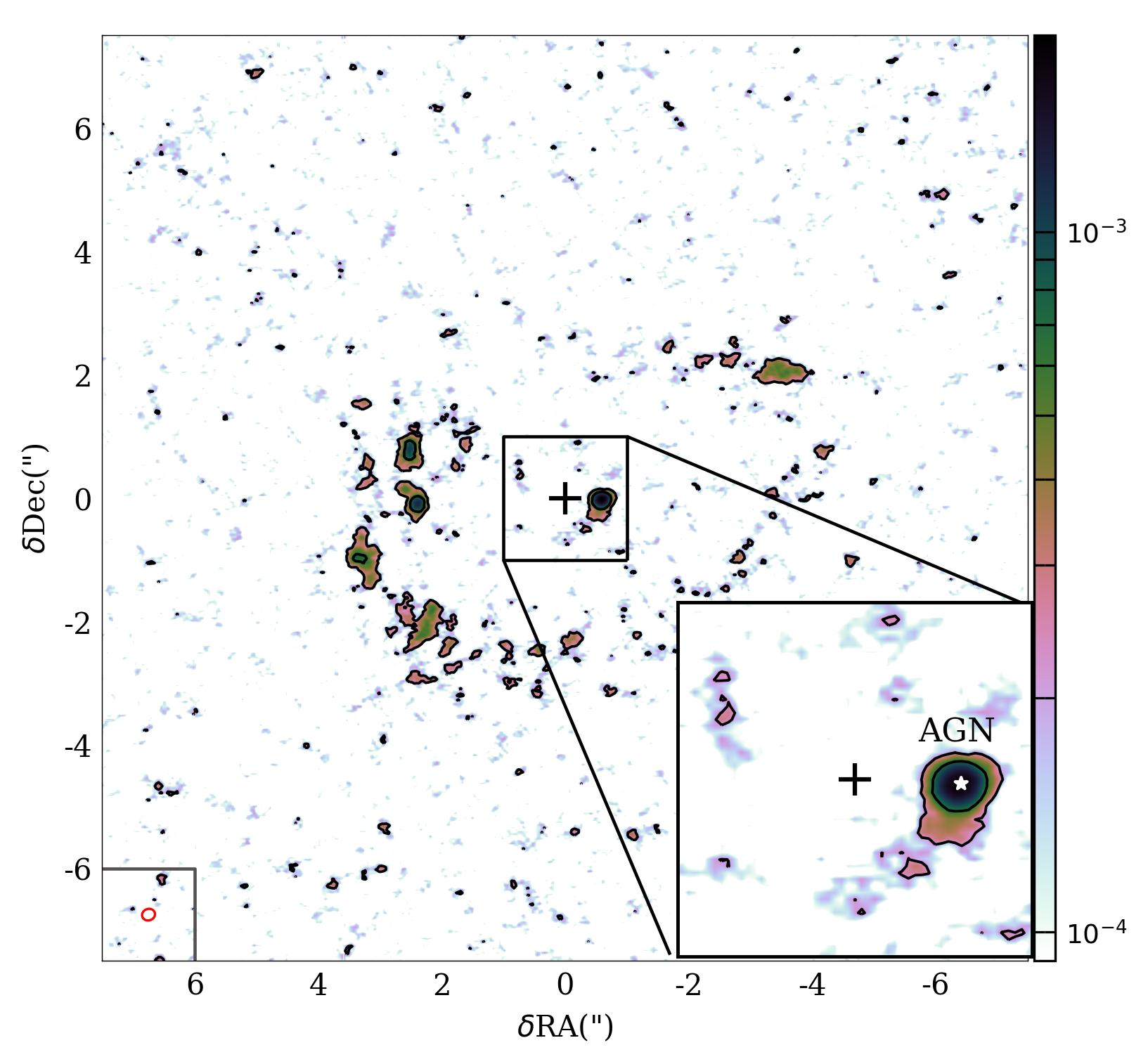}}
\caption{Continuum emission at 0.87\,mm. The central 15\arcsec$\rm \times$15\arcsec\ are shown in the main panel, the zoom-in of the central 2\arcsec$\rm \times$2\arcsec\ is shown in the right bottom corner. The cross indicates the phase center, while the white star the new center adopted in Table~\ref{tab:prop}. The colour scale is in Jy/beam. The beam size (0.21\arcsec$\times$0.19\arcsec) is shown in the red ellipse in the bottom left corner. }
\label{fig:cont}
\end{figure}

\section{Results}\label{results}

\subsection{Continuum emission}\label{cont_emission}

\begin{figure*}[h!]
\centering
\includegraphics[width=17cm]{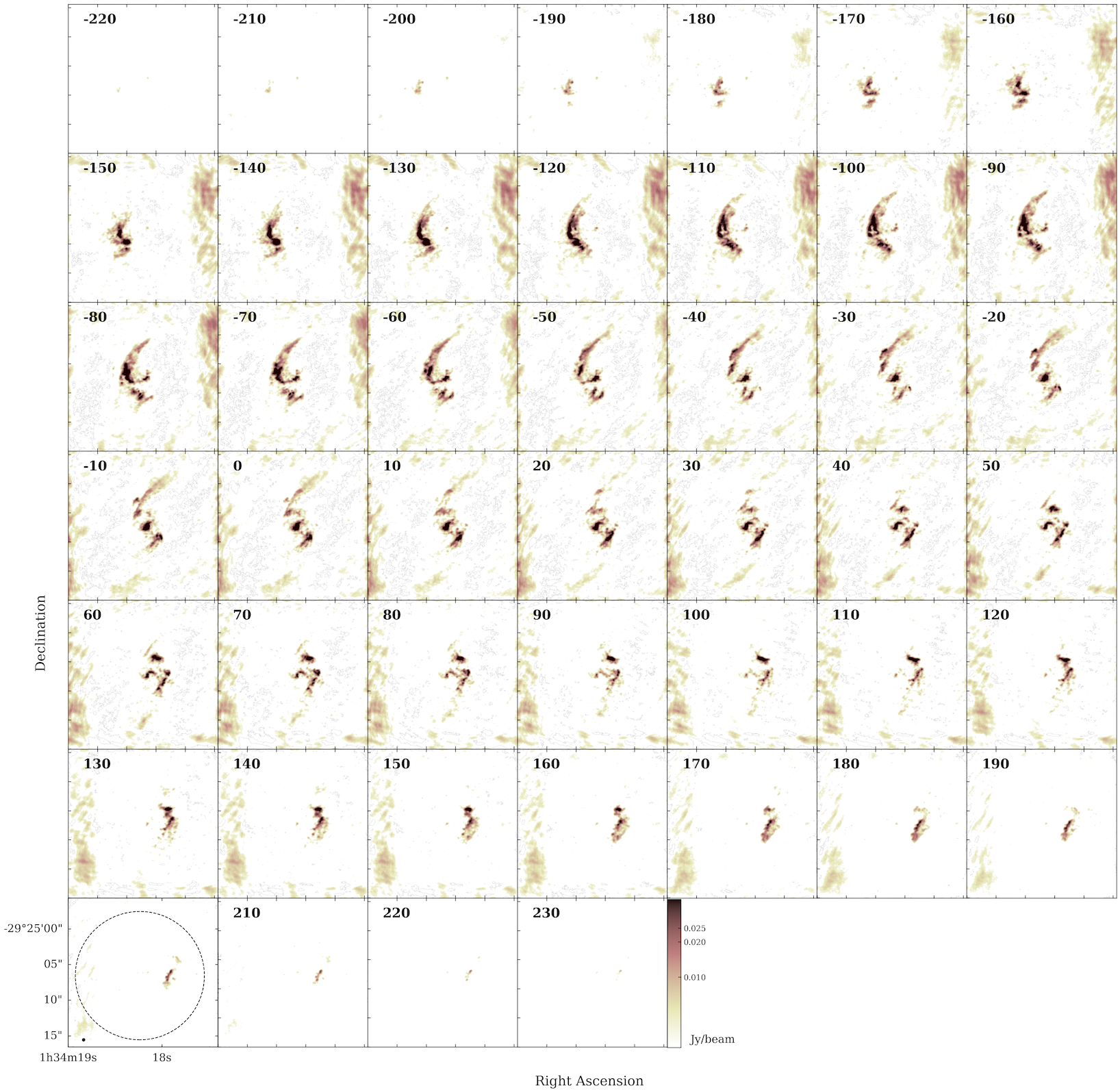}
\caption{Channel maps of CO(3-2) emission in the center of NGC\,613 using the combined observations from ALMA Cycles 3 and 4. We present the channels maps from  from -220 (top left) to +230\,km/s (bottom right) relative to the $v_{sys}=$1471\,km$\rm s^{-1}$, in steps of 10\,km/s. Each of the 46 square boxes is 21\arcsec$\times$21\arcsec ~in size, while the primary beam is 18\arcsec ~in diameter and it is indicated in the dashed circle on the bottom left panel. The synthesized beam (0\farcs21$\times$0\farcs19, PA=-69$^\circ$) is shown in the black ellipse in the bottom left corner. The center of the maps is the phase centre of the interferometric observations given in Table 1. The colour scale is in power stretch, ranging between 2 and 38\,mJy/beam.}
\label{fig:chans}
\end{figure*}

Previous ALMA band 3 and 7 observations by \citet{miy17} detect continuum emission from both the circumnuclear disk (CND) and the star-forming ring (250$\rm<r<$340\,pc). At 95\,GHz with a $\sim$0\farcs6 resolution, they find a continuum jet with a PA=20$^\circ$, which corresponds to the 4.9\,GHz and 14.9\,GHz jets \citep{hum92}, close to the minor axis of the ring. In the nucleus, the negative spectral index, $\alpha\sim$-0.6, is compatible with synchrotron emission, with a small fraction of free-free, while the index $\alpha\sim-0.2$ along the star-forming ring can be from free-free emission \citep{miy17}.

At our resolution of 0\farcs2 ($\sim 17$\,pc), the central continuum is resolved at 350\,GHz, with some compact emission along the star-forming ring, as display in Figure~\ref{fig:cont}. The $\sim$2.2\,mJy peak emission is detected at 25$\sigma$ significance. We determine the peak of the continuum emission by fitting a circular Gaussian source in the $uv$-plane using the GILDAS \textsc{uv\_fit} task. The fitted results in the central continuum emission for the flux is 2.4$\pm$0.1\,mJy and for the RA and DEC relative do the phase center are $\Delta$RA=-0\farcs587 and $\Delta$DEC=-0\farcs03219, with a relative uncertainty of $\pm$0\farcs003. These values are listed as reference of the new adopted center and AGN position in Table~\ref{tab:prop}. Within the error bar, the AGN position is consistent with the positions derived by \citet{miy17} and from X-ray observations \citep{liu05}.

%The total continuum emission integrated over our FoV amounts to 12.5 mJy, which is consistent with the extrapolation of  the far-infrared SED  from previous measurements, if account is taken for the missing flux and the continuum extent: the total flux for the whole galaxy should be of the order of 1 Jy. Most of the emission must come from  outside the ALMA FoV, and it is likely that the interferometer misses  a significant part of the emission, more so than for the line where the  velocity splitting reduces the extent of the emission, in each channel.

\subsection{Molecular gas distribution and morphology}

\begin{figure}
\resizebox{\hsize}{!}{\includegraphics{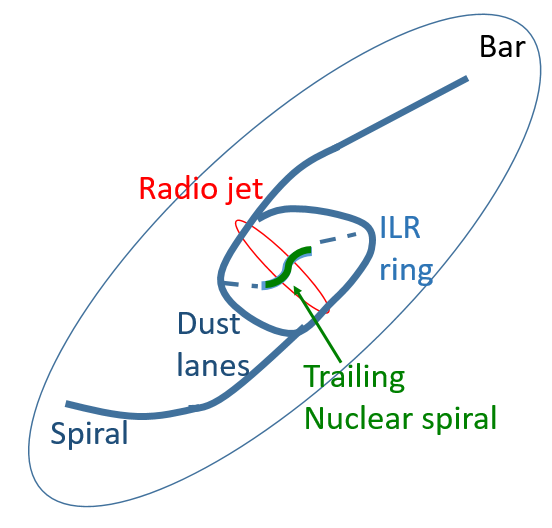}}
\caption{Sketch of the main morphological features observed in the CO(3-2) emission in NGC\,613. The orientation of the dust lanes follows the large scale bar. The filaments are shown in dashed lines between the ILR star-forming nuclear ring and the nuclear trailing spiral. The radio jet is also shown for comparison.}
\label{fig:sketch}
\end{figure}

\begin{figure*}
\centering
\includegraphics[width=17cm]{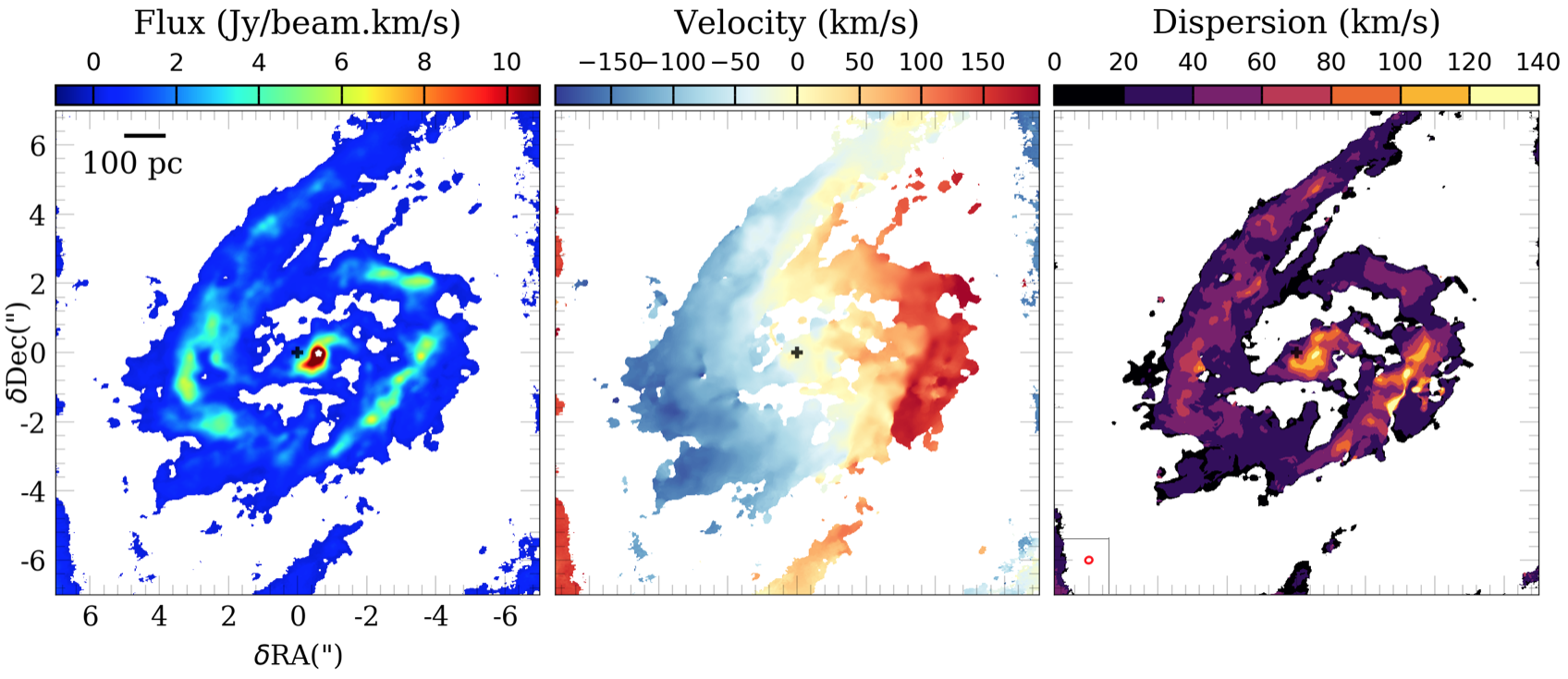}
\caption{CO(3-2) moments maps of NGC\,613 in the central 14\arcsec ($\sim$1.2\,kpc) FoV.  We show the integrated intensity map (0$th$ moment, \textit{left}), intensity-weighted velocity  map (first-moment, \textit{middle}) and the intensity-weighted velocity dispersion map (second-moment, \textit{right}). The black cross indicates the phase center and the white star the new adopted AGN position (in Table~\ref{tab:prop}). The synthesised beam of 0\farcs21$\times$0\farcs19 is shown in red in the bottom left corner of the second-moment map.}
\label{fig:mom_613}
\end{figure*}

Figure~\ref{fig:chans} displays the CO(3-2) channel maps, with a velocity range of 450\,km/s and a velocity resolution of 10\,km/s. The channels show evidence of a regular velocity field in a ring at a radius $\sim$3.5\arcsec (300\,pc), with two winding arms structure coming from the NW and SE directions. These spiral arms coincide with the beginning of the dust lanes along the bar seen in the HST/F814W image (Figure~\ref{fig:hst}): they are the contact points between the tangent dust lanes and the ring. 

At small radii, there is a spiral structure in the center channels ($\pm$100\,km/s) that can be more clearly seen in Figure~\ref{fig:mom_613}. The 2-arm nuclear gas spiral at $\rm r\lesssim$100\,pc is trailing toward the center, and it will be discussed further in Section~\ref{torque}. Additionally, the velocity distribution is perturbed for channels between $\pm$50-100\,km/s for $\rm r\sim$150\,pc, and the morphology shows evidence of a filamentary structure connecting the ring and the nuclear spiral. The main morphological features are show in the sketch of the galaxy in Figure~\ref{fig:sketch}.

We constructed the moment maps of the CO(3-2) line clipping the emission at $<$5$\rm\sigma_{rms}$.  The integrated intensity (zero-moment) map in the left panel of Figure~\ref{fig:mom_613} shows that the CO emission follows the $\sim$300\,pc star-forming circumnuclear ring. \cite{miy17} have mapped the ring in CO(1-0) and CO(3-2) with ALMA at 0\farcs7 and 0\farcs4 respectively, and found a clumpy ring, globally regular, but with spots of active and efficient star formation. In our maps at higher resolution, we find that the molecular emission in the ring is clumpy and incomplete, and  coincides with the same star forming clumps observed in Br\,$\gamma$ in the near-infrared (NIR) \citep[see][and Sect. \ref{nir}]{fal14}. The CO(3-2) emission peak of 25\,Jy.km/s/beam corresponds to the AGN position reported in Table~\ref{tab:prop}. Within the CND, there is a clear trailing 2-arm spiral structure. The ring reveals two breaks into two winding spiral arms, at NW and SE. 

We superposed in Figure~\ref{fig:hst} the CO(3-2) contours onto the \textit{HST} maps in the F814W filter\footnote{The HST image was aligned to the ALMA astrometry, the peak emission in the HST image was recentered to the AGN position in Table~\ref{tab:prop}.}. It shows a remarkable similarity in morphology, the molecular ring seen in the CO emission coincides with the dusty nuclear ring in the HST image, and the winding arms are the beginning of the characteristic dust lanes along the bar. At the very center ($\lesssim$100\,pc) though, the stellar and molecular morphologies are dissimilar.

In the middle panel of Figure~\ref{fig:mom_613}, the intensity-weighted velocity (first-moment) map shows a clear rotation pattern in the galaxy plane, with velocities peaking between $\sim \pm$200\,km/s from the systemic velocity ($v_{sys}$1481\,km/s, see discussion below). The velocity distribution and morphology in the central 200\,pc are more perturbed, due to the filamentary streams and the nuclear spiral. The NW winding arm is mostly blueshifted and the SE redshifted, indicating that rotation and possibly gas pile-up is taking place; this will be discussed further in Section~\ref{torque}. 

Close to the AGN, the velocity dispersion is high ($\sigma\sim$130\,km/s), as displayed in the right panel of Figure~\ref{fig:mom_613} (second-moment map). In the nuclear spiral the velocity dispersion ranges from 70$\sim$120\,km/s and the average dispersion along the ring is $\sim$40\,km/s, with more elevated values in the clumpy regions. Furthermore, we can distinguish a disturbance in the overdense region in the west part of the ring, with an increased dispersion $\gtrsim$150\,km/s. This region also corresponds to an enhanced spot observed in [Fe\textsc{ii}] with SINFONI, suggesting a strongly shocked medium (see also Sect.~\ref{nir}).

\subsection{CO luminosity and H$\rm_2$ mass}

\begin{figure}
\resizebox{\hsize}{!}{\includegraphics{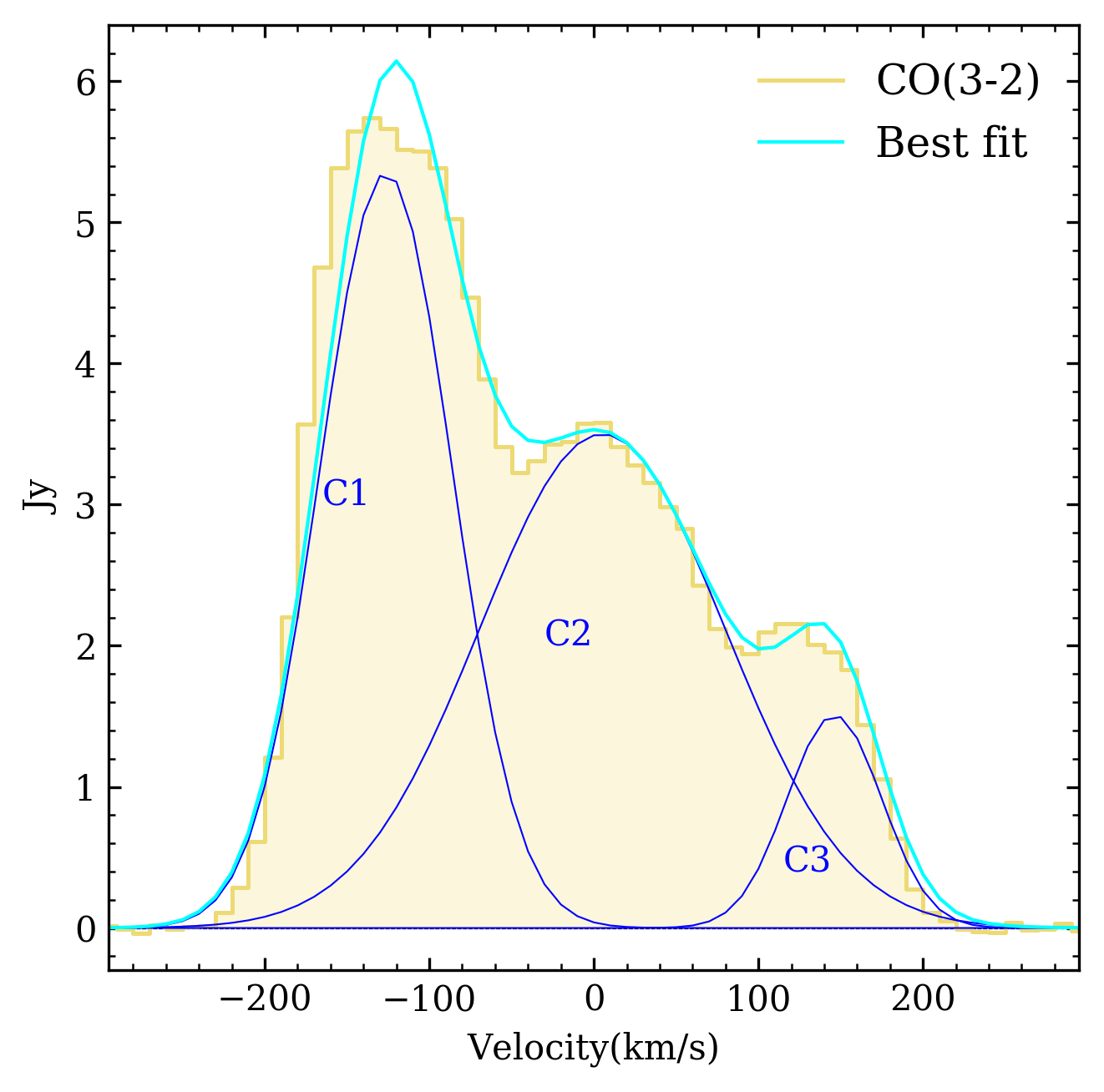}}
\caption{We show the total CO(3-2) emission line profile integrated over the observed map, with a FoV of 18", after correction for primary beam attenuation. The light blue line is the result of the Gaussian fit with three velocity components (in dark blue); see Table~\ref{tab:flux}.}
\label{fig:tot}
\end{figure}

The mean intensity map is plotted in Fig.~\ref{fig:mom_613} (left). Since the galaxy is more extended than the primary beam, it is difficult to quantify the missing flux. We compare it to the central spectrum obtained with the 15-m single dish obtained with the Swedish-ESO Submillimeter Telescope (SEST) in CO(1-0) and CO(2-1) over a 43\arcsec~ and 22\arcsec~ FoV, respectively. In Figure \ref{fig:tot}, we display the total CO(3-2) spectrum integrated over the 18\arcsec FoV. Towards the central position, \citet{bajaja95} found a CO(2-1) spectrum peaking at T$_A^*$= 200\,mK with  FWHM=300\,km/s, yielding a total integrated flux of 1504\,Jy km/s, in a beam of 22\arcsec. Their beam is very similar to our FoV of 18\arcsec. The flux comparison is relevant, since our FoV encompasses the entire nuclear ring, and the emission in this nuclear region corresponds to the strongest surface density at different wavelengths \citep{com10,ho11,li11}, as already discussed by \citet{combes1566}.

We assume a ratio of $\rm r_{31}=T_{3-2}/T_{1-0}$ of 0.82, typical for Seyfert galaxies \citep{mao10} and a ratio $\rm r_{21}=T_{2-1}/T_{1-0}$ compatible with $1$, within the error bars. The latter is derived from the SEST CO(2-1)/CO(1-0) observations in the galaxy center, by convolving the beam of CO(2-1) to 43\arcsec, implying a higher CO excitation at the center of NGC\,613. This is expected for thermalized excitation and a dense molecular medium. In that case, the CO(3-2) flux should be higher than the CO(2-1), as we could presume the flux $\rm S_{\nu}\propto\nu^2$ in the Rayleigh-Jeans approximation, for gas at temperature larger than 25K and density larger than 10$^4$ cm$^{-3}$). Using these values, the expected CO(3-2) intensity is $\sim$2266\,Jy.km/s in a 22\arcsec\ beam. When integrated over the spectral range (FWHM$\sim$250km/s), the integrated emission in our ALMA FoV of 18\arcsec\, shown in Figure~\ref{fig:tot}, is 1307\,Jy.km/s. Therefore, we should expect some missing flux by a factor up to $\sim$40-50\%, taking into account the uncertainties of the $r_{31}$ and $r_{21}$ ratios. Hence, we find a molecular mass of 5.6$\times$10$^8$M$_\odot$ in our FoV, assuming  thermally excited gas, and a Milky-Way like CO-to-H$_2$ conversion factor, of 2$\times$10$^{20}$cm$^{-2}$/(K.km/s)  \citep[e.g.][]{bol13}. 

In comparison, the SEST CO(1-0) observations of \citet{bajaja95} give a total molecular mass of 3.8$\times$10$^9$\,M$_\odot$ integrated over 33 pointing positions covering $\sim$120\arcsec along the galaxy, and in the central 43\arcsec\ beam yields to $\rm M_{H2}\approx$1.5$\times$10$^9$\,M$_\odot$. In the 22\arcsec\ beam, the SEST CO(2-1) spectrum, together with the CO(2-1)/CO(1-0) ratio of $1$, gives a mass of 1.2$\times$10$^9$\,M$_\odot$.  

The total CO(3-2) emission line profile integrated over the observed map (FoV of 18\arcsec) is shown in Figure~\ref{fig:tot}. We decomposed the spectrum in three components, C1, C2 and C3, and the results of the Gaussian fits for each component are displayed in Table~\ref{tab:flux}. The total flux is $\rm S_{CO(3-2)}=1307\pm121$\,Jy.km/s.

\begin{table}
\caption{Line fluxes}              % title of Table
\label{tab:flux}      
\centering                                      
\begin{tabular}{l c c c c}         
\hline\hline  
\noalign{\smallskip}
Line & $\rm S_{CO(3-2)}$ & V & FWHM  & $\rm {S_{peak}}^a$ \\  
        & (Jy.km/s)  &   (km/s) &  (km/s) & (Jy) \\
\hline   
\noalign{\smallskip}                               
%CO(3-2) & 1307.2 $\pm$ 6.1  &  -58.9$\pm$0.4  &  256.0$\pm$0.9  &  4.8 \\
C1 & 541.1$\pm$54.7  &  -126.3$\pm$1.7  &  95.0$\pm$3.6  &  5.4 \\
C2 & 654.0$\pm$90.7  &  5.2$\pm$5.0  &  175.5$\pm$20.8  &  3.5 \\
C3 & 109.1$\pm$27.7  &  146.2$\pm$2.8  &  68.0$\pm$8.4  &  1.5 \\
\hline
\noalign{\smallskip}
%HCN(4-3) & 25.7$\pm$0.5 & -5.4$\pm$1.4 & 149.1$\pm$3.9 & 0.16 \\
%HCO$\rm^+$(4-3) & 19.1$\pm$0.3 & -21.0$\pm$1.8 & 241.1$\pm$ 4.2 & 0.07 \\
%CS(7-6) & 2.0$\pm$0.08 & -9.1$\pm$2.5 & 135.8$\pm$6.5 & 0.014 \\
\end{tabular}
\tablefoot{Results of the Gaussian fits for the 3 velocity components (C1, C2 and C3), shown in Fig.~\ref{fig:tot}. \\
\tablefoottext{a}{Peak flux}
}
\end{table}

\subsection{CO(3-2) kinematics}\label{kin}

\begin{figure}
\resizebox{\hsize}{!}{\includegraphics{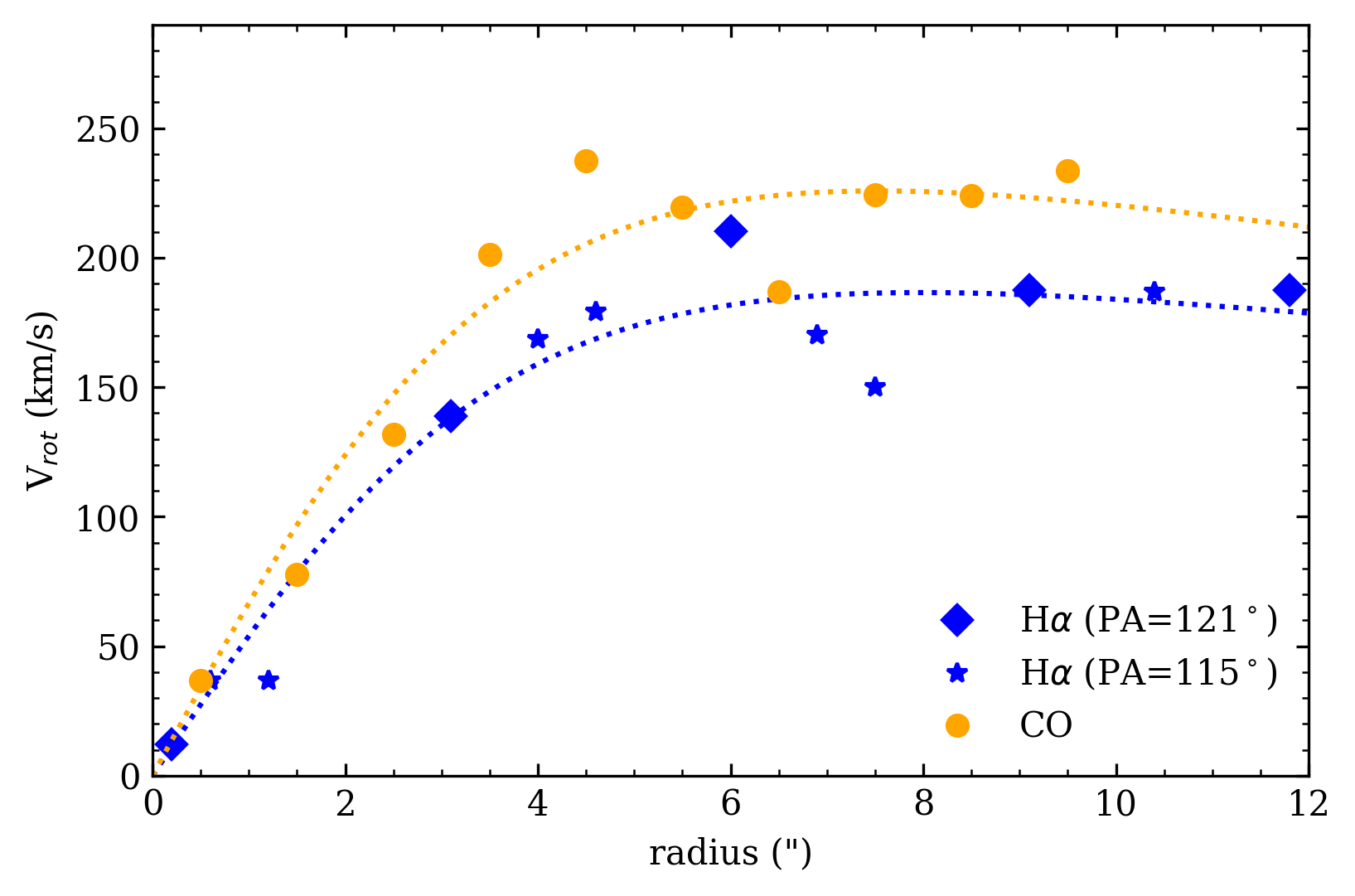}}
\caption{Rotation curve of NGC\,613. The orange circles represents the CO kinematics from our ALMA observations and the blue diamonds and stars are the H$\alpha$ measurements by \citet{bur64}, for a PA of 121$^\circ$ and 115$^\circ$, respectively. The dotted lines are the best fit assuming the gas on circular orbits in a plane, $v_{c}=Ar/(r^2 +c^2)^{p/2}$.}
\label{fig:vrot}
\end{figure}

\begin{figure*}
\centering
\includegraphics[width=17cm]{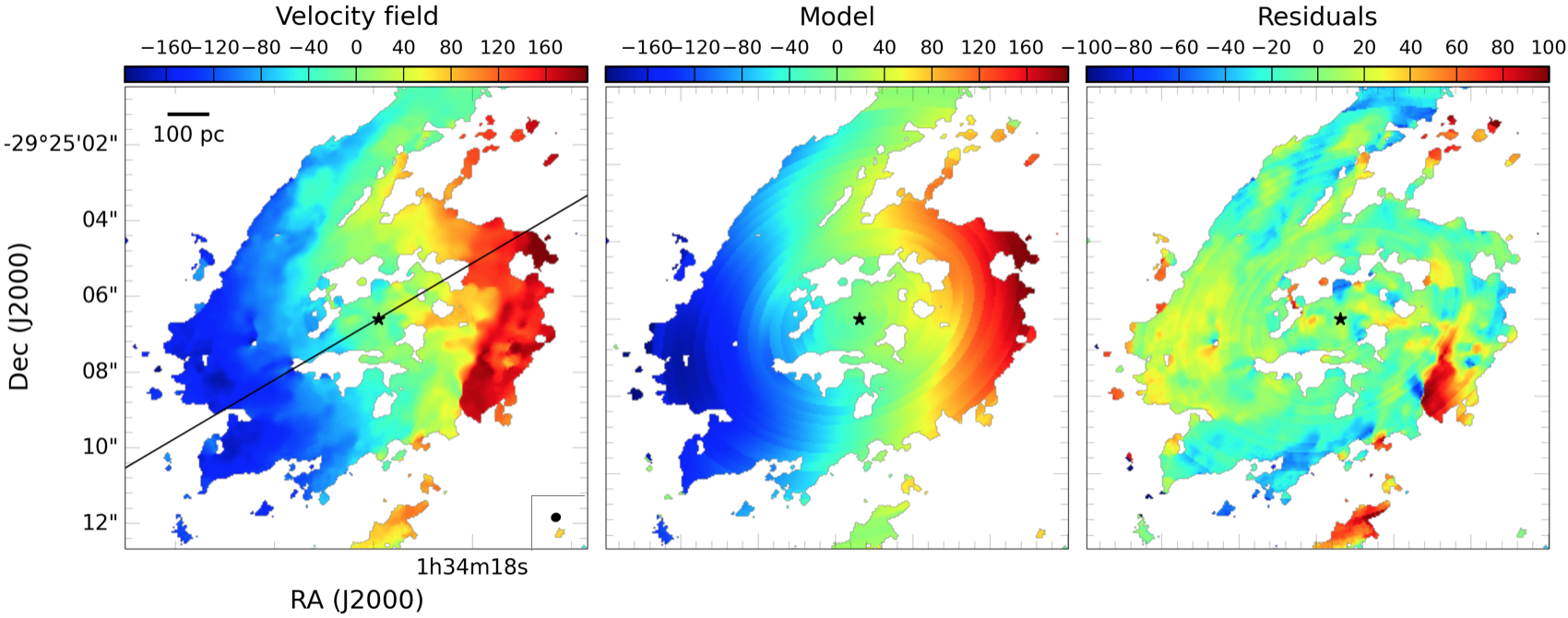}
\caption{{\it Left:} Velocity map of NGC\,613 clipped at $>$5$\sigma$. In the {\it middle} panel the best fit model using the tilted-ring approach for the \citet{ber91} model and the residuals, after subtracting the model from the data, are shown on the {\it right}. The line in the \textit{left} indicates the average estimated major axis PA=120$^\circ$ and the black stars are the central positions adopted as the AGN position.  The synthesised beam is shown in black ellipse in the bottom right corner of the first panel.}
\label{fig:vmod}
\end{figure*}

In Figure~\ref{fig:vrot} we show the rotation velocities deduced from the CO kinematics and the H$\alpha$ rotation curve taken from \citet{bur64}. As already pointed out, the dominant feature in velocity field of the molecular gas appears to be due to circular rotation in the disk (middle panel of Fig. \ref{fig:mom_613}), and is consistent with the H$\alpha$ kinematics.  We find a good agreement with the PA and inclination from optical studies, in the range of PA=111$\sim$124$^\circ$  and i=36$\sim$47$^\circ$ \citep{bur64,bla81,vau91}, and therefore we adopted the values of PA=120$^\circ$ and i=41$^\circ$, listed in Table~\ref{tab:prop}. We assume a simple model for the rotation curve, proposed by \citet{ber91}, assuming the gas is on circular orbits in a plane, $v_{c}=Ar/(r^2 +c^2)^{p/2}$, where $A$, $c$, and $p$ are parameters of the model, and for $p=1$ the velocity curve is asymptotically flat and $p=3/2$ the system has a finite total mass, therefore we expect $1\leq p \leq 3/2$. Figure~\ref{fig:vrot} shows the result of fitting the radially averaged velocities to this model;
the dotted lines represent the best fit for circular velocity, $v_c$. 

We can refine the above fit of the radially averaged velocities by fitting the entire velocity field
with the same model by \citet{ber91}. The observed radial velocity at a position $\rm (R,\Psi)$ on the plane of the sky can be described as: 
\begin{eqnarray}\label{eq:ber}
v(R,\Psi)= v_{sys} +~~~~~~~~~~~~~~~~~~~~~~~~~~~~~~~~~~~~~~~~~~~~~~~~~~~~~~~~~~~~~~~~~~~~ \\
\frac{A R cos(\Psi -\Psi_0)sin(\theta)cos^p(\theta)}{\left \{ R^2 \left[ sin^2(\Psi -\Psi_0)+ cos^2(\theta)cos^2(\Psi -\Psi_0) \right] +c^2 cos^2(\theta) \right \} ^{p/2} } \nonumber
\end{eqnarray}
where $\theta$ is the inclination of the disk (with $\theta$ = 0 for a face-on disk), $\Psi_0$ is the position angle of the line of nodes, $v_{sys}$ is the systemic velocity and $R$ is the radius. We used the tilted-ring model \citep{rog74}, which consists in dividing the velocity field in concentric rings in radii $\Delta r$ and each ring is allowed to have an arbitrary $v_c$, $i$ and PA. For each radius, we can independently fit the parameters of Equation~\ref{eq:ber} to the observed velocity field. 

We show the results of the tilted-ring fitting to the velocity map in Figure~\ref{fig:vmod}. We adopted a $\Delta r$=0\farcs3, that corresponds to the deprojected resolution of our observations in the galaxy plane. In the right panel, we display the residuals after subtracting the \citet{ber91} model to the velocity field. As can be seen, the model represents quite well the observed velocity field, with no significant amplitudes in the residuals along the galaxy disk, except in the west part of the ring, where there is an important contribution from non-circular motions. This region coincides with the contact point between the ring and the SE winding arm, and it is also probably perturbed by shocks, as suggested by the high velocity dispersions in the molecular gas and an enhancement in the [Fe\textsc{ii}] emission (Fig.~\ref{fig:sinf}). This non-circular motions along the minor axis and on the winding spiral indicate streaming motions associated with inflow, if we assume trailing perturbations and that the north side is the far side of the galaxy, as the NW arm is blueshifted and the SE arm is redshifted. 

\begin{figure}
\resizebox{\hsize}{!}{\includegraphics{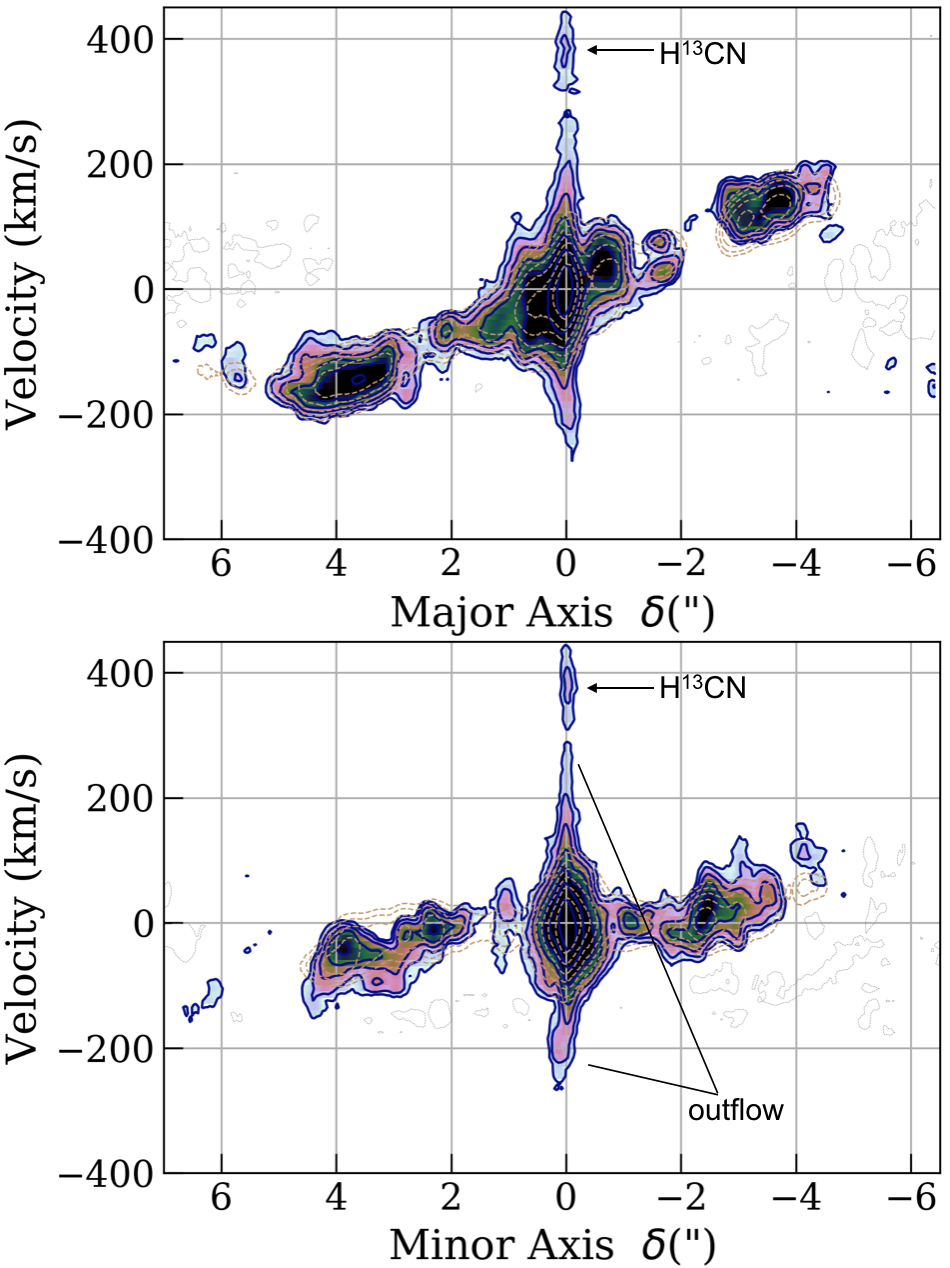}}
\caption{Position-velocity diagrams of NGC\,613 for the CO(3-2) emission along the major axis at a PA=120$^\circ$ (\textit{top panel}) and minor axis (PA=210$^\circ$, \textit{bottom}). We used a 0\farcs2 slit width. The dashed lines are the best fit from  \textsuperscript{3D}BAROLO. The emission around $v\sim$400 km/s corresponds to the isotope H$\rm ^{13}$CN(4-3).}
\label{fig:pv}
\end{figure}

An additional method was used to derive the CO kinematics using the ``3D-Based Analysis of Rotating Objects from Line Observations" ( \textsuperscript{3D}BAROLO) software by \citet{barolo}. \textsuperscript{3D}BAROLO performs a 3D tilted-ring modeling of the emission line data-cubes to derive the parameters that better describe the kinematics of the data. We ran \textsuperscript{3D}BAROLO on the CO(3-2) data-cube in order to investigate non circular motions, since the code allows us to infer radial velocities in the fit of the rotation curves. We have performed several tests running the code, varying the fixed and free parameters of the disk model, and the results that better reproduce the observed velocity field are found when fixing the PA, inclination and central position to 120$^\circ$, 41$^\circ$ and AGN position, respectively. The best fit reveals radial components of order $\rm v_{rad}\sim$20\,km/s in the nuclear region and $\rm v_{rad}\sim$20--100\,km/s from the end of the circumnuclear ring at 4\arcsec up to r$\sim$7\arcsec, corresponding to the streaming motions of the winding arms. Overall, we find that in the case of high-resolution ALMA observations, the results using \textsuperscript{3D}BAROLO are quite coherent with the 2D approach described above in the text. 

The position-velocity diagrams (PVDs) along the major (PA=120$^\circ$) and minor axis (210$^\circ$) of NGC\,613 are shown in Fig.~\ref{fig:pv}. The best fit obtained with \textsuperscript{3D}BAROLO including radial velocity components is shown in dashed lines. We notice a highly skewed kinematics in the center, with velocities $\gtrsim$200\,km/s, that cannot be described only by co-planar circular motions in the galaxy disk. Since the velocities in the very centre ($\lesssim$0.5\arcsec) strongly deviate from the rotation curve pattern, we believe this is a signature of an outflow emanating from the AGN. We discuss the detection and properties of the molecular outflow in Section~\ref{outflow}.

\begin{figure*}
\centering
\includegraphics[width=17cm]{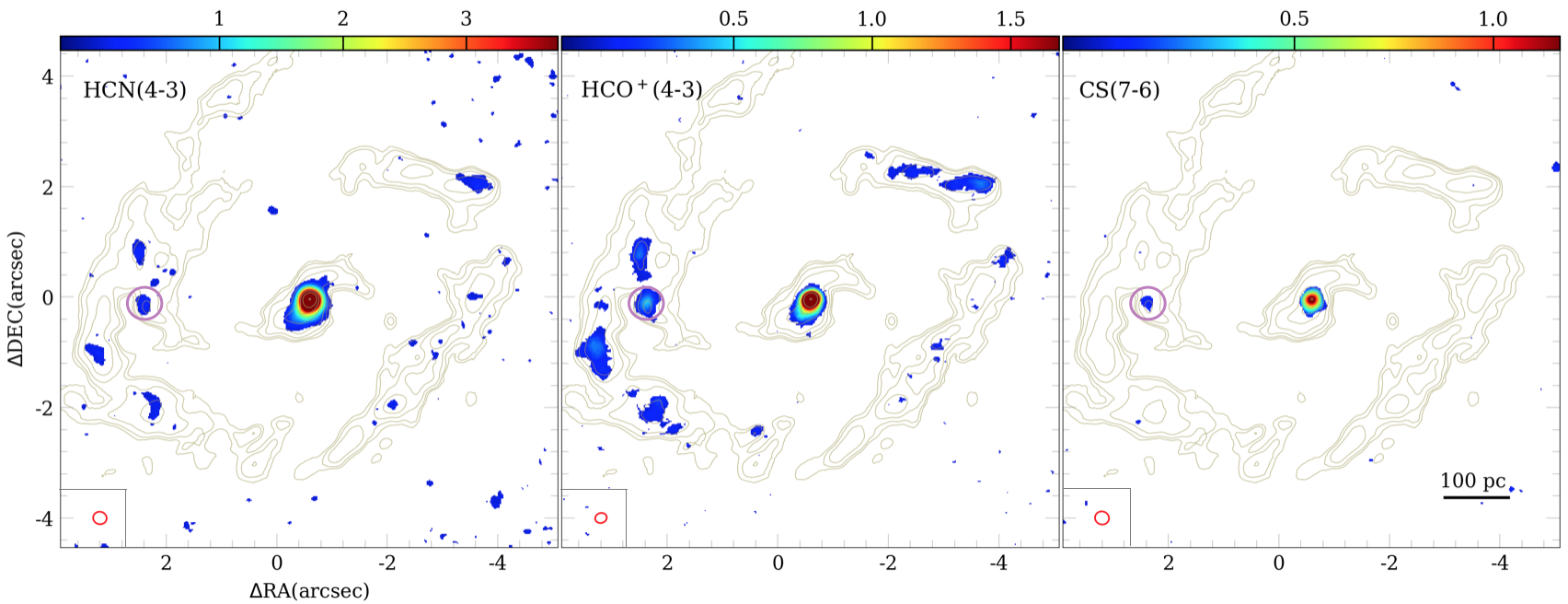}
\caption{The integrated intensity maps of the dense gas tracers HCN(4-3), HCO$\rm ^+$(4-3) and CS(7-6) from the left to right panels, with the CO contours overplotted. HCN and HCO$\rm ^+$ present clumpy emissions along a radius of $\sim$3\arcsec, coinciding with the star-forming ring. All the dense gas tracers are detected in the very center, as nuclear compact decoupled disks, or molecular tori as discussed in Paper \textsc{I}. The purple circles indicate the position of a clump detected in the HCN, HCO$\rm ^+$ and CS emission and used to calculate the ratios in the sub-millimeter diagram in Fig.~\ref{fig:submm}. The synthesised beam sizes are shown in red in the bottom left corner of each panel.}
\label{fig:dense}
\end{figure*}

\begin{figure*}
\centering
\includegraphics[width=17cm]{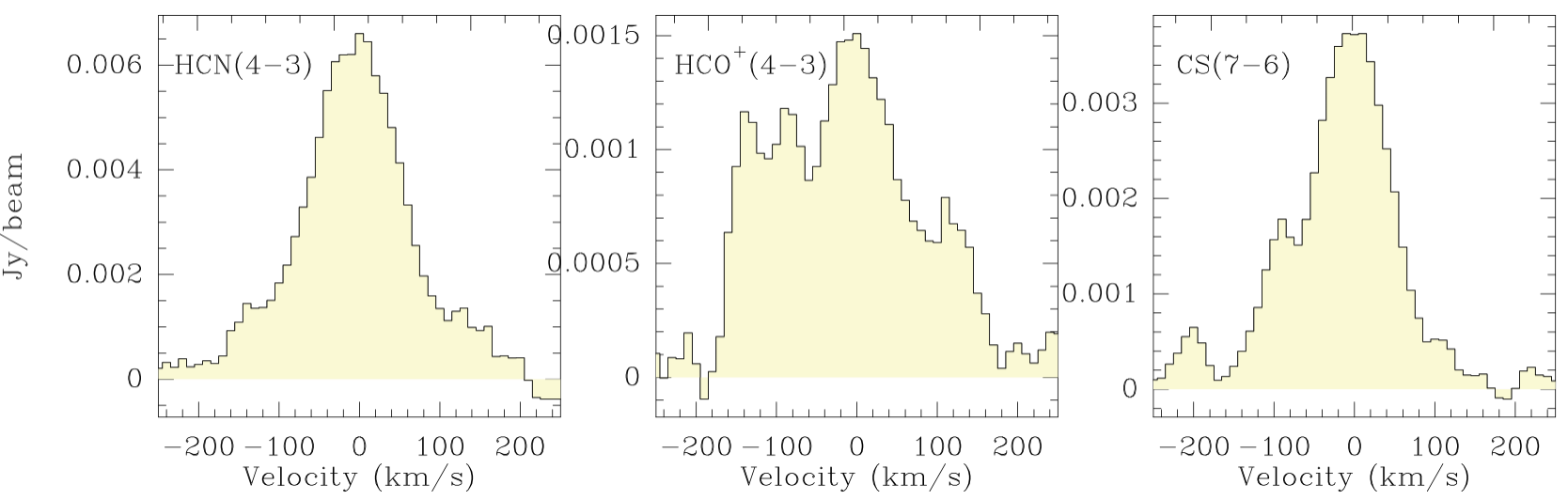}
\caption{We show the HCN(4-3), HCO$^{\rm +}$(4-3) and CS(7-6) (from \textit{left} to \textit{right}) emission line profiles integrated in along our FoV, using the mask from the zero-moment map in Fig.~\ref{fig:dense}. The integrated values fitting a Gaussian are listed in Table~\ref{tab:flux}.}
\label{fig:dense_spec}
\end{figure*}

\subsection{Dense gas: HCO$^+$, HCN and CS emission}\label{dense}

In ALMA band 3, dense gas is detected in various lines HCN(1-0), HCO$^+$(1-0) and CS(2-1), while SiO(2-1) was marginally detected at the edges of the radio jets, probably indicating the existence of shock regions related to the jets, as reported by \citet{miy17}. Along with CO(3-2), our band 7 observations detected the HCO$^+$(4-3), HCN(4-3) and CS(7-6) emission lines. The corresponding maps are displayed in Figure \ref{fig:dense}, and the integrated spectra in Figure \ref{fig:dense_spec}. CS(7-6) is mainly detected in the center of the galaxy, while for HCO$^+$(4-3) and HCN(4-3), we detect stronger emission in the center but also some clumps along the star-forming ring. The dense gas tracers detected in the very center are interpreted to be a molecular torus, as discussed in Paper I.
The HCN line is about twice brighter than the HCO$^+$ line in the nuclear region, which is the typical value expected for AGN \citep[e.g.,][]{kohno05,krips08,ima16}. The detection of these lines reveals the presence of dense gas, since the critical densities of the HCO$\rm ^+(4-3)$, HCN(4-3) and CS(7-6) transitions are 2.6$\times$10$^6$, 1.4$\times$10$^7$ and 3.4$\times$10$^6$\,cm$^{-3}$, respectively. 

\begin{figure}
\resizebox{\hsize}{!}{\includegraphics{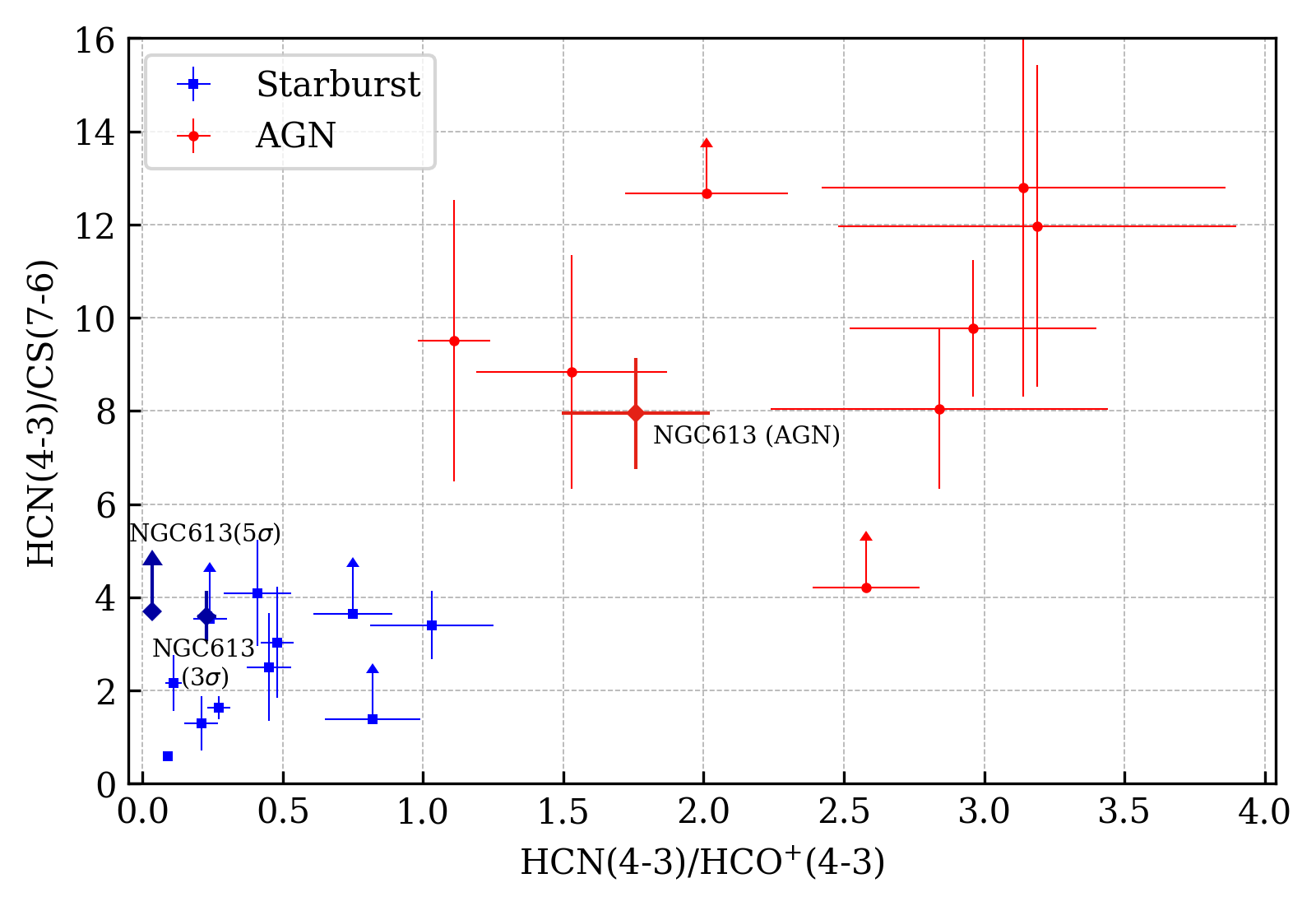}}
\caption{The sub-millimeter-HCN diagram proposed by \citet{izumi16} for the high-resolution observations (spatial resolution $<$500\,pc) using the line intensity ratios R$_{\rm HCN/HCO^{+}}$ and R$_{\rm HCN/CS}$. The red circles represent the AGN and the blue squares indicate the SB galaxies. We include the line ratios of NGC\,613 (diamonds) measured in the CND, called here as "AGN" and in a clump detected $\sim$250\,pc north-est of central position in all the dense tracers shown in Fig.~\ref{fig:dense}}
\label{fig:submm}
\end{figure}

\begin{figure*}
\centering
\includegraphics[width=17cm]{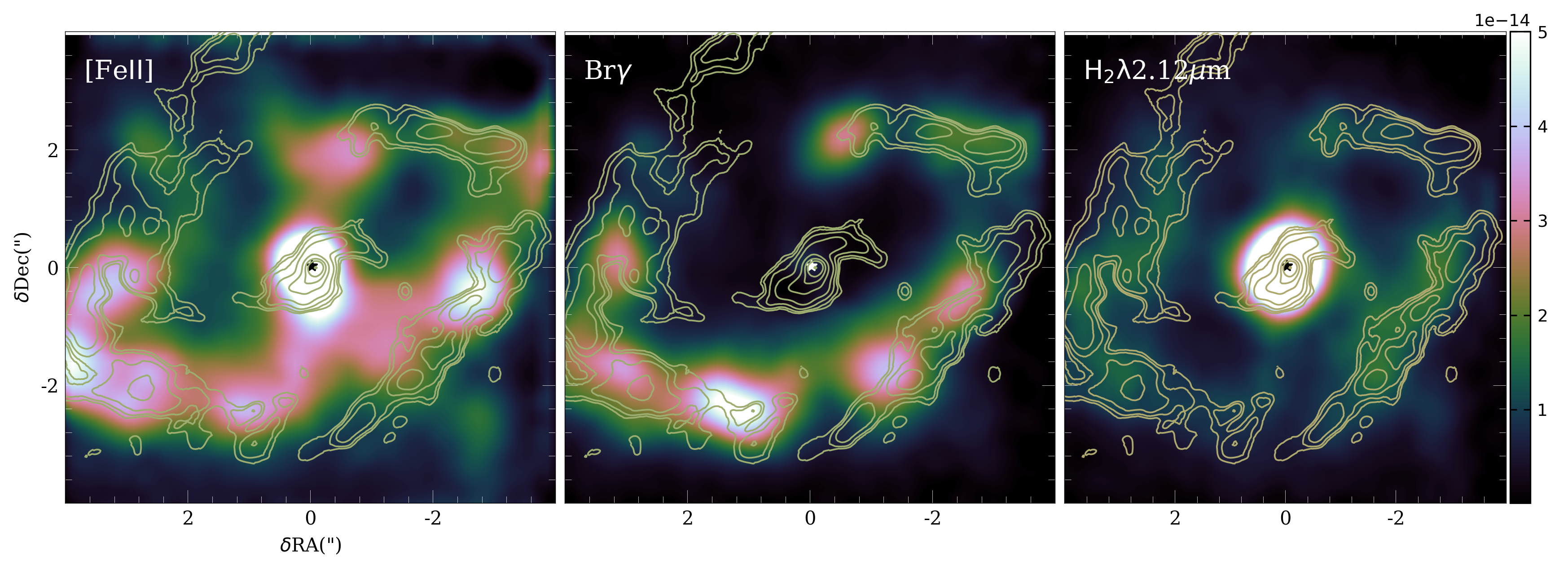}
\caption{Comparison between the CO(3-2) emission, shown in contours, with the [Fe\textsc{ii}] (\textit{left}), Br$\gamma$ (\textit{middle}), and H$\rm_{2}\lambda 2.12\mu m$ (\textit{right}) emission in the 8\arcsec$\times$8\arcsec FoV of the SINFONI observations \citep{fal14}. The black/white star represents the AGN position listed in Table~\ref{tab:prop}.}
\label{fig:sinf}
\end{figure*}

As already mentioned, NGC\,613 exhibits different ionization mechanisms. Recently, spatially resolved Baldwin, Phillips \& Terlevich (BPT) diagrams \citep{bpt81}, can isolate the contributions from star formation, shock excitation and AGN activity using optical line ratios, as studied with the Siding Spring Observatory Wide-Field Spectrograph (WiFeS) by \citet{davies17} and European Southern Observatory Multi-unit Spectroscopic Explorer (MUSE) in \citet{gado19}. Likewise, in the sub-millimeter domain, multi-line observations of higher J transitions are fundamental to derive the chemical conditions of the molecular gas and the heating mechanisms \citep{viti14,ima18}. One useful extinction-free energy diagnostic tool in the centers of galaxies is the sub-millimeter-HCN diagram (Fig.~\ref{fig:submm}) proposed by \citet{izumi16}. The diagram uses the HCN(4-3)/HCO$^+$(4-3) and HCN(4-3)/CS(7-6) ratios to distinguish the dominant energy source exciting the molecular gas in galaxies, whether by AGN (XDR or X-ray dominated region) or star formation (PDR or photodissociation region).  They suggest enhanced integrated intensity ratios in circumnuclear molecular gas around AGN compared to those in starburst galaxies (sub-millimeter HCN-enhancement). 

Thanks to our high-resolution ALMA observations, we are able to disentangle the emission coming from the nuclear region within the spiral trailing feature observed in the central $\sim$100\,pc and the contribution from a star-forming clump observed in all three molecular tracers at $\sim$250\,pc from the nucleus. The clump is indicated as a circle in Figure~\ref{fig:dense}. We measured the line intensity ratios R$_{\rm HCN/HCO^{+}}$ and R$_{\rm HCN/CS}$ in these two regions (CND/AGN, for a central aperture of 0\farcs3 and clump, listed in Table~\ref{tab:ratio}). For a 5$\sigma$ threshold, the clump is barely detected in CS(7-6) emission, and we present the line ratios for a 3$\sigma$ clump detection and a upper limit for the R$_{\rm HCN/CS}$.

The line ratios are plotted on the sub-millimeter diagram of \citet{izumi16}, as displayed in Figure~\ref{fig:submm}. We find that the CND region lies in the AGN-dominated part of the diagram, while the clump in the star forming ring of NGC\,613 is indeed dominated by star formation. Ultimately, we do find that the nuclear region of NGC\,613 presents line ratios that indicate excitation conditions typical of XDRs in the vicinity of AGN.

\begin{table}
\caption{Line ratios: R$_{\rm HCN/HCO^{+}}$ and R$_{\rm HCN/CS}$}          
\label{tab:ratio}      
\centering                                      
\begin{tabular}{c c c}         
\hline\hline        
\noalign{\smallskip}               
Region & R$_{\rm HCN/HCO^+}$   & R$_{\rm HCN/CS}$  \\  
\hline       
\noalign{\smallskip}                           
 AGN & 1.76$\pm$0.26  &   7.95$\pm$1.19   \\
 Clump (3$\sigma$)& 0.23$\pm$0.03 & 3.6$\pm$0.5 \\
 Clump (5$\sigma$) & 0.034$\pm$0.005 &  $>$3.7$\pm$0.6 \\
\hline        
\end{tabular}
\tablefoot{Ratios of the HCN, HCO$\rm^+$ ans CS lines in the CND region and in the clump shown in Fig.~\ref{fig:dense}. \\
}
\end{table}

\subsection{Comparison to the warm molecular and ionised gas}\label{nir}

In order to compare the CO(3-2) morphology with the ionised material and the warm molecular gas, we superposed the CO contours onto the NIR maps of [Fe\textsc{ii}], Br$\gamma$ and H$\rm_{2}\lambda 2.12\mu m$ presented in \citet{fal14}, shown in Figure~\ref{fig:sinf}. There is a remarkable resemblance between the ionised and warm molecular gas to the CO emission along the star forming ring.  The position and ages of the hot spots in the ring suggest a \textit{``pearls on a string”} scenario of evolution of star formation as proposed by \citet{boker08}. In this scenario, star formation only occurs in particular overdense regions and the young clusters move along the ring, following the gas movement, and meanwhile age, resulting in an age gradient along the ring. The expected sequence of star formation was indeed observed in the southern part of the ring: the hottest stars are found near the contact point of the dust lanes, and then fewer hot stars are found along the ring \citep[see Fig.~8 of][]{boker08}.

At the center, the nuclear spiral corresponds to the massive reservoir of the bright warm H$\rm_2$ and [Fe\textsc{ii}], while contrasting with the weak emission in Br$\gamma$. Indeed, as discussed by \citet{fal14}, the high [Fe\textsc{ii}]/Br$\gamma$ ratio in the center indicates that excitation is dominated by shocks and photoionization in the nucleus, and follows the correlation between the strength of the [Fe\textsc{ii}] and 6\,cm radio emission in Seyfert galaxies \citep{forbes93}. 

The high values of the ratio [Fe\textsc{ii}]/Br$\gamma$=17.7 in the nucleus of NGC\,613, are not typical of starburst galaxies, where we expect the ratio in the range 0.5-2 \citep{col15} and the [Fe\textsc{ii}] emission originates in supernova-driven shocks. Indeed, the large ratio in the nucleus is similar to those found in AGN, indicating that the most likely mechanism for the production of [Fe\textsc{ii}] emission in  is shock excitation from the radio jets and/or supernova remnants (SNRs), typical for Seyfert galaxies \citep{ardila04}. In fact, X-ray emission, which is dominant in Seyferts, can penetrate deeply into atomic gas and create extended partly ionized regions where [Fe\textsc{ii}] can be formed. Models presented by \citet{alo97} show that X-rays are able to explain [Fe\textsc{ii}]/Br$\gamma$ ratios up to $\sim$20, in agreement with the values observed in NGC\,613 nucleus. 

\begin{figure}
\resizebox{\hsize}{!}{\includegraphics{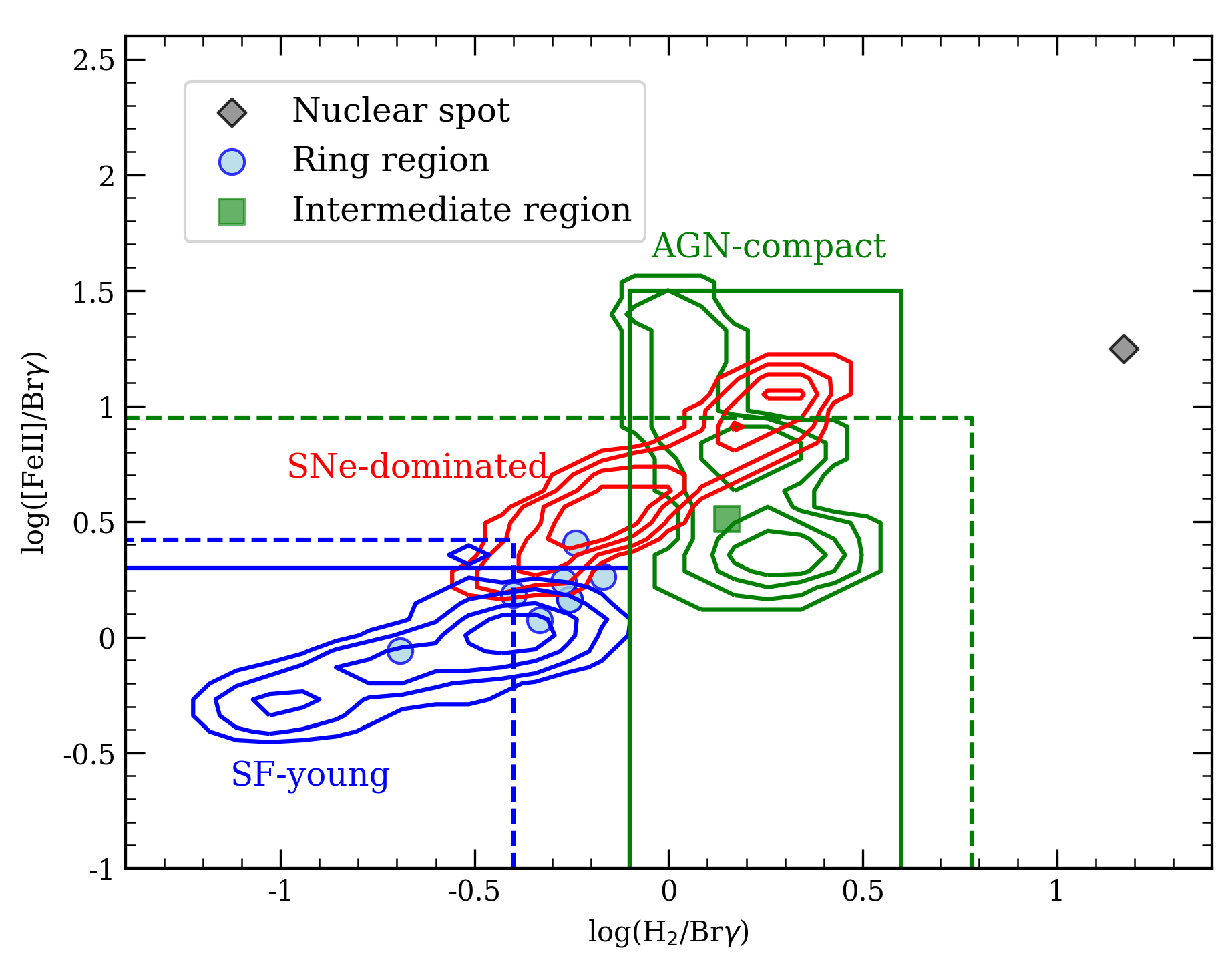}}
\caption{Diagnostic diagram with the NIR emission line ratios H$_2$(1- 0)S(1)$\lambda$2.122\,$\mu$m/Br$\gamma$ and \fe2\,$\lambda$1.644$\mu$m/Br$\gamma$. Apertures in the circumnuclear ring are shown in blue circles and the aperture from the nuclear region in gray diamond. Contours denote regions for young star formation, supernovae, and compact AGN, while solid lines denote upper limits for young star formation and AGN, both derived from integral-field spectroscopy data \citep{col15}. Dashed lines denote upper limits for star formation and AGN derived from slit spectroscopy \citep{rogerio13}.}
\label{fig:nir}
\end{figure}

\citet{col15} developed a 2D diagnostic diagram using integral field spectrograph spectrograph data to characterize line emitting regions. The diagnostic uses the line ratios H$_2$(1- 0)S(1)$\lambda$2.122\,$\mu$m/Br$\gamma$ and\fe2\,$\lambda$1.644$\mu$m/Br$\gamma$. By this method, they found that young star-forming regions, older supernova dominated regions, and the compact AGN dominated region occupy different areas in the line-ratio space. We show the NIR diagnostic diagram proposed by \citet{col15} in Figure~\ref{fig:nir}. We use the line emission listed in Tables~1 and 2 of \citet{fal14} for different apertures: one in the nucleus, 7 along the circumnuclear ring (spots 1 to 7) and 1 aperture between the ring and the nucleus \citep[``spot 8'', see Fig.1 of][]{fal14}.  Placing the line ratios for the different apertures in the diagrams, we note that all spots in the circumnuclear ring are located in the regions of young star formation or SNe-dominated stellar populations. The spot in the intermediate region has higher line ratios, and it is shifted into the compact AGN region. 
On the other hand, the nucleus of NGC\,613 takes its place in the shock ionized LINER regime, where we expect H$_2$/Br$\gamma$ ratios larger than 6 \citep{mazza13,rogerio13}. As suggested by \citet{fal14}, the enhancement of the H$_2$ emission in the nucleus is possibly due to the interaction with the radio jet. The high value of H$_2$/Br$\gamma$ measured in the nucleus of NGC\,613 is consistent with a LLAGN Seyfert/LINER composite, which is strongly influenced from shock heating. 

\section{The molecular outflow}\label{outflow}

\subsection{The CO(3-2) kinematics of the outflow}\label{kine}

The PVDs along the major and minor axes of the galaxy shown in Fig.~\ref{fig:pv} illustrate the highly skewed kinematics in the central part of NGC\,613. The nuclear region contains gas reaching velocities up to $\sim\pm$300\,km/s in projection, which is much higher than the rest of the nuclear disk gas, as we can notice in the middle panel of Fig.~\ref{fig:mom_613}.

\begin{figure}
\resizebox{\hsize}{!}{\includegraphics{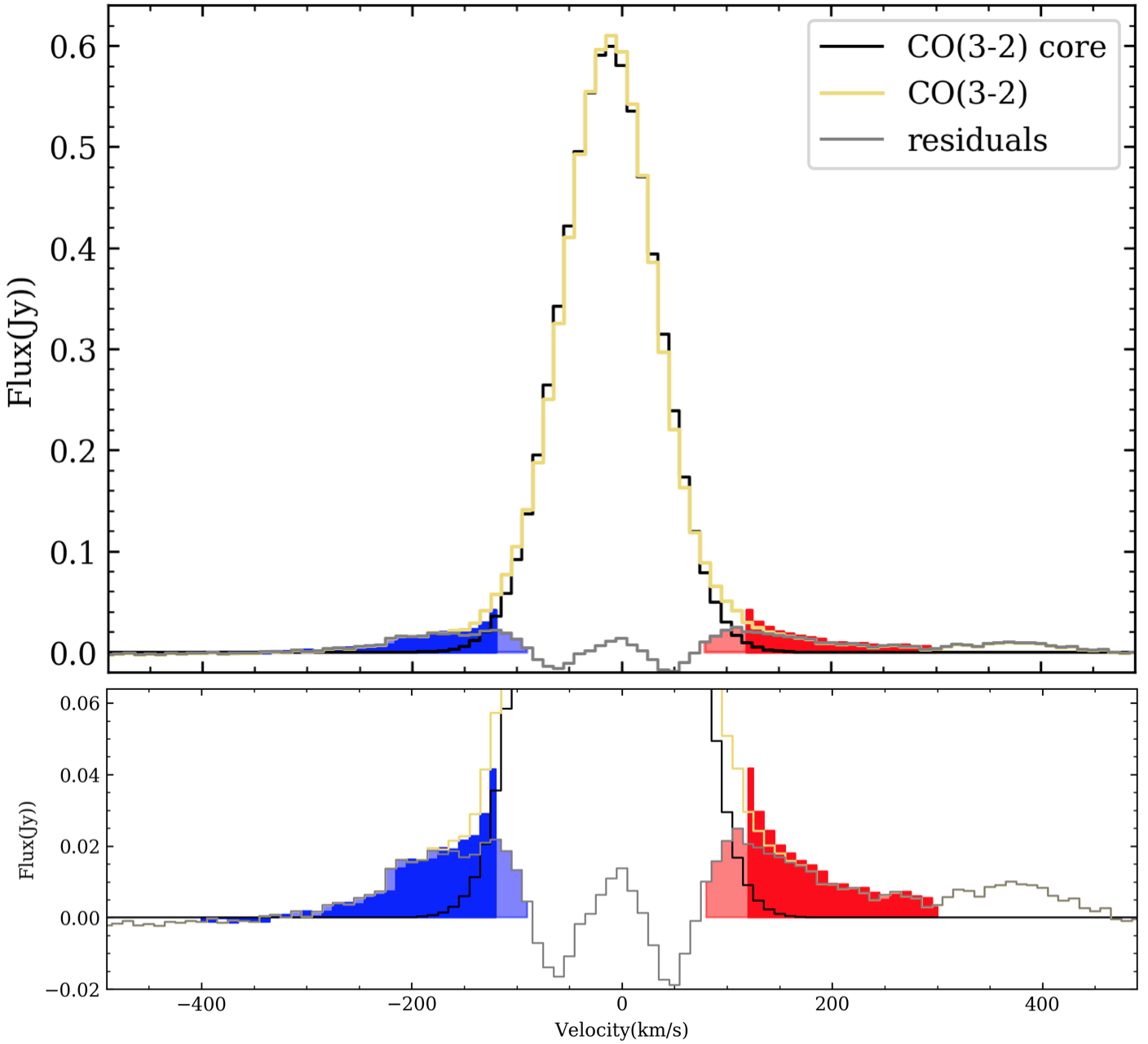}}
\caption{In the top panel we show the nuclear CO(3-2) spectrum in yellow, extracted in a 0\farcs28 region. We fit a Gaussian that takes into account the main disk/core contribution to the CO emission and subtract to the observed spectrum (residuals in gray). The regions  considered to the computation of the molecular outflow properties in the blue and red wings are show in color. \textit{Bottom:} zoom view of the blue (-400 to-120\,km/s) and red (120 to 300\,km/s) wings. The red wing is integrated only up to +300\,km/s to avoid the contamination from the H$\rm ^{13}$CN emission at $\sim$400\,km/s.}
\label{fig:outfit}
\end{figure}

An examination of the individual spectra at pixels around the AGN position reveals conspicuous blue and redshifted wings in all the spectra within a radius of $r\sim$0.3\arcsec. 
These broad line wings are characteristic of gas ejection out of equilibrium, in this case, an indication of a molecular outflow. The integrated spectrum extracted in a central circular aperture of 0.28\arcsec~($\sim$23\,pc) is shown in Figure~\ref{fig:outfit}. We can notice a bump in the emission around $\rm v\sim$400\,km/s, which corresponds to the isotope H$^{13}$CN(4-3).

One might question whether the outflow signature could arise by a central mass able to induce such a high rotation in the centre. If we assume that these high-velocity components are within the SoI, in a radius equivalent to our nuclear aperture of $\sim$0.3\arcsec =25\,pc, or about r=38\,pc in the galaxy plane, a massive black hole located in the centre should have a mass of at least $M_{BH}$=$\rm v^2$R/G for the rotational velocity in the galaxy plane $v$=460\,km/s (corresponding to $\pm$300\,km/s projected velocities), or M$_{BH}$=1.9$\times$10$^9$\,M$_\odot$. This value is about 2 orders of magnitude higher than the values reported in the literature, e.g. M$_{BH}$=7.4$\times$10$^6$\,M$_\odot$ derived by \citet{davis14} using the spiral pitch angle and M$_{BH}$=4$\times$10$^7$\,M$_\odot$ using the central stellar velocity dispersion by \citet{bosch16}. In Paper I, we have derived a black hole mass of $M_{BH}=3.7\times10^{7}M_\odot$ within the SoI of 50\,pc, and for all the NUGA sample the values derived tend to follow the pseudo-bulge region in the M$_{BH}-\sigma$ plane \citep{ho14}. 

Furthermore, if these high-velocity features were due to the rotation, they should not be observed along the minor axis of the galaxy; however, the PVDs in Figure~\ref{fig:pv} clearly show a gradient of the high-velocity emission along the minor axis that cannot be explain as due to co-planar rotation. Could these features be signature of inflowing gas? We cannot exclude a priori that the high-velocity components are due to a coplanar inflow, since the blueshifted gas is observed in the north and the redshifted gas is the south. If the far side of the galaxy is the north side, this gradient could be explained as an inflow. However, the  deprojected velocities onto the galaxy disk would be about $v\sim$300/sin(i)$\sim$460\,km/s. The order of magnitude of the co-planar inflow would be too large, and therefore we can exclude this hypothesis. 

\begin{figure*}
\centering
\includegraphics[width=17cm]{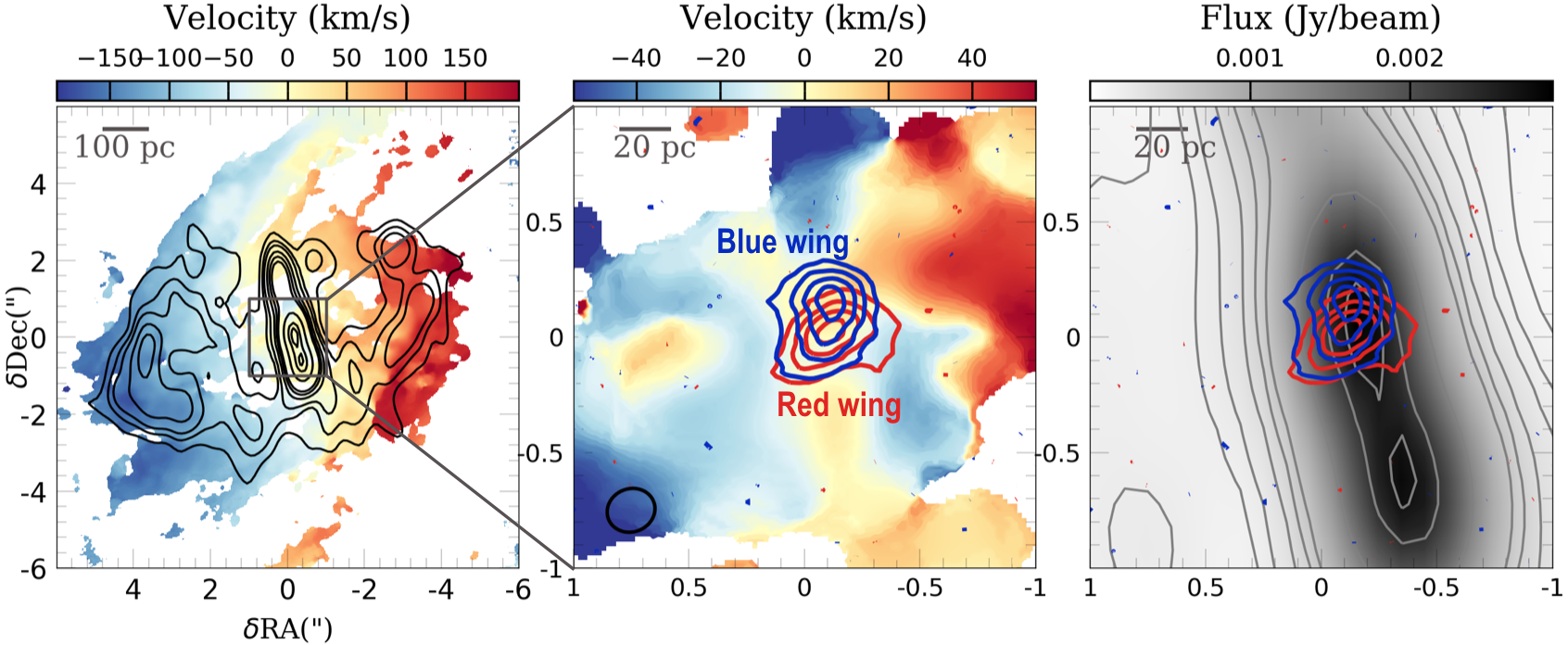}
\caption{\textit{Left}: The velocity distribution of the CO(3-2) emission with the VLA radio contours at 4.86\,GHz over-plotted. \textit{Middle:} a 2\arcsec $\times$2\arcsec zoom of the velocity distribution and the emission contours of the blue and red wings shown in Figure~\ref{fig:outfit}. The synthesised beam is shown in black in the bottom left corner. \textit{Right:} The grayscale and contours of the 4.86\,GHz radio emission and the blue and red wings contours. The direction of molecular outflow is aligned with the radio jet, and spatially corresponds to the central blob of the radio map.}
\label{fig:wings}
\end{figure*}

Additionally, in Figure~\ref{fig:wings} we show the contours for the blue and red wing emission, and we see that the contours overlap and the blue component arises in the northern part of the nucleus and the red in the south. Due to the small size of the outflow, we can barely resolve each wing contribution, however, there is an indication that they do not follow the rotation pattern of the mean velocity field. The direction of velocities are also opposite to what is found in the molecular torus (cf Paper I).

We compare the molecular to the radio emission observed with the Karl G. Jansky Very Large Array (VLA) at 4.86\,GHz \citep{hum92} in Figure~\ref{fig:wings}. The VLA radio jet emission is shown in the right panel with the molecular outflow  blue and red wings emission  contours overlaid. The molecular outflow emission coincides with the central blob of the radio jet and appears to be aligned with the orientation of the radio jet at PA=12$^\circ$.

In order to derive conservative values for the flux related to the outflow in the broad emission, instead of fitting one Gaussian component for the \textit{``core''} and one broad component for the outflow (including some low-velocity emission that might not be associated with the outflowing material), we have taken two different approaches. First, we fit a Gaussian to the nuclear spectrum to take into account the contribution of the main rotating disk (black line in Fig.~\ref{fig:outfit}). After subtracting the \textit{core} contribution to the CO spectrum, we have the residuals, as shown by the gray line. We then integrated the contribution of the blue and red regions in the residual spectrum. The results of the fit of the main core and blue and red wings in the residuals are listed in Table~\ref{tab:outfit}. The derived molecular mass corresponding to the main disk component inside the radius of 23\,pc is $\rm4.8\times10^{7}M_\odot$, which agrees with the mass of the 14\,pc molecular torus of $\rm 3.9\times10^{7}M_\odot$ found in Paper I (using the same conversion factor and $\rm r_{31}$ as in Paper I).

\begin{table}
\caption{Gaussian fit in the nuclear spectrum}              % title of Table
\label{tab:outfit}   
\begin{tabular}{lcccc}
\hline \hline
\noalign{\smallskip}
Component & $\rm S_{CO}$  & Position & FWHM & $\rm {S_{peak}}$\\
  & (Jy.km/s) & (km/s) & (km/s) & (mJy) \\
\hline
\noalign{\smallskip}
Core$\rm^a$ & 68.5 $\pm$ 1.1  &  -11.8 $\pm$ 0.4  &  107.2 $\pm$ 0.9  &  600 \\
%H$\rm^{13}$CN & 1.1 $\pm$ 1.6  &  368.6 $\pm$ 133.2  &  117.1 $\pm$ 301.6  &  9.0 \\
\end{tabular}
\begin{tabular}{p{1.4cm}p{2cm}p{3cm}p{1.2cm}}
\noalign{\smallskip}
\hline
\noalign{\smallskip}
\multicolumn{1}{c}{} & \multicolumn{1}{c}{$\rm S_{CO}$} & \multicolumn{1}{c}{Velocity Range} &   \multicolumn{1}{c}{$\rm {S_{peak}}$}\\
\multicolumn{1}{c}{} & \multicolumn{1}{c}{(Jy.km/s)} & \multicolumn{1}{c}{(km/s)}  & \multicolumn{1}{c}{(mJy)} \\
   \hline
\noalign{\smallskip}
\multicolumn{4}{c}{\textsc{Integrated Residual Wings}}\\
\noalign{\smallskip}
\hline
\noalign{\smallskip}
Blue & 2.63 $\pm$ 0.8  &  [-400,-90]    &  21.8 \\
Red & 2.62 $\pm$ 0.9  &  [+80,+300]  &  24.9 \\
\noalign{\smallskip}
\hline
\noalign{\smallskip}
\multicolumn{4}{c}{\textsc{Integrated Red and Blue Maps}}\\
\noalign{\smallskip}
\hline
\noalign{\smallskip}
Blue & 2.3$\pm$ 0.8  &  [-400,120]    &  41.5$\rm^b$ \\
Red & 2.3 $\pm$ 0.9  &  [+120,+300]  &  41.7$\rm^b$ \\
\hline
\end{tabular}
\tablefoot{\tablefoottext{a}{Results of the Gaussian fit for the main disk contribution of the nuclear spectrum extracted in a $r$=0.28\arcsec aperture shown in Fig.~\ref{fig:outfit}}. \\
\tablefoottext{b}{In this case we just assumed that $\rm {S_{peak}}$ is the maximum flux in the nuclear spectrum within the selected velocity range of the blue and red wings.} 
}
\end{table}

The second approach consists in creating moment maps only taking the velocity channels from -400 to -120\,km/s for the blue component and from +120\,km/s to 300\,km/s for the red wing. We try to avoid the contribution from the H$\rm ^{13}$CN by limiting the velocity channels up to 300\,km/s. We display the blue and red wings contour maps in Figure~\ref{fig:wings}, superposed to the intensity weighted velocity map of the original CO map, which represents the mean velocity pattern. The integrated fluxes corresponding to velocity intervals of the red and blue maps are also listed in Table~\ref{tab:outfit}. The peak temperatures of the blue and red wings represent $\sim$5\% of the peak of the main core component, however they are still detected with a 50$\sigma$ significance in the case of the residuals wings. 

The maximal velocities of the red and blue components are up to about $\pm$300\,km/s in projection ($\sim$460\,km/s if in the galaxy plane, $i$=41$^\circ$). Given their location near the nucleus, we tentatively interpret these high-velocity features as the two sides of an outflow. Globally, these features represent as much as $\lesssim$8\% of the total molecular emission in the nuclear ring region. 

\subsection{CO-to-$H_2$ conversion in the nuclear region of NGC613}

From the integrated flux $\rm S_{CO}\Delta V$(Jy km/s) listed in Table~\ref{tab:outfit}, we can derive the molecular mass involved in the outflow using the equation from \citet{sol05}:
$$
{L_{CO}}^\prime(K.km/s/pc^2) = 3.25\times10^7 \frac{S_{CO}\Delta V}{1+z}  \left( \frac{D_{L}}{\nu_{rest}} \right)^2 
$$
where $\rm \nu_{rest}=345.796$\,GHz, and $\rm D_L$ is the luminosity distance in Mpc. The molecular mass, including helium, is then derived from $\rm M(H_{2}) =\alpha_{CO}{L_{CO}}^\prime r_{13}$ \citep{tacco13}. This implies a molecular mass of $M_{out}$=1.9-2.2$\times$10$^6$M$_\odot$. This mass was obtained using $r_{31}$=0.82, a luminosity distance of 17.2\,Mpc and the standard Galactic CO-to-H$_2$, conversion factor \citep[$\rm\alpha_{CO,MW}=4.36\,M_\odot(K.km/s pc^2)^{-1}$][]{dame01,bol13}. However, this mass could be an upper limit if the flow is made of more diffuse optically thin gas.

The standard $\rm \alpha_{CO}=4.36$ for the Milky Way is the recommended value to use in the inner disk of galaxies. However, several observational works \citep[e.g.,][]{israel09a,israel09b,sand13} found that in the center of galaxies (R$\lesssim$1\,Kpc) the conversion $X_{CO}$ can be a factor up to 3-10 times lower than $X_{CO,MW}$. As pointed out by \citet{bol13}, the recommended value to be applied in galaxy centers is $\alpha_{CO,cen}\sim\frac{1}{4}\alpha_{CO,MW}$, with a 0.3\,dex uncertainty. In our case, the masses involved in the outflow would be four times lower, i.e., in the range of $\rm M_{out}=4.8-5.5\times10^5 M_\odot$, providing a more conservative estimative of the mass. 

The assumption of a smaller $\rm \alpha_{CO}$ has been already discussed in the literature. In the case of the outflow in NGC\,1068, the $\frac{1}{4}\alpha_{CO,MW}$ factor was also assumed \citep{santi1068}, in agreement with LVG analysis of the CO line ratios in the central region of this galaxy \citep{usero04}. Another example is the molecular outflow detected in M\,51, where the authors assumed $\rm \alpha_{CO}=\frac{1}{2}\alpha_{CO,MW}$ \citep{que16a,mat07}.  In a study of molecular gas excitation
in the jet-driven winds of IC\,5063, \citet{kal16} found that the outflowing molecular gas is partly optically thin, implying a $\rm \alpha_{CO}$ one order of magnitude smaller than the Galactic.

Most molecular outflows are detected in ultra-luminous infrared galaxies (ULIRGs) \citep{cicone14}. Therefore, the CO-to-H$_2$ conversion factor usually assumed in the literature is $\rm\alpha_{CO}=0.8$, a factor $\sim$5 times lower than the Milky Way factor $\alpha_{CO,MW}$. Since NGC\,613 has a rather moderate infrared luminosity ($L_{IR}=3\times10^{10}L_\odot$), there is no reason a priori to adopt the lower factor of applied to ULIRGs. In fact, we would like to highlight that uncertainties in $\alpha_{\rm CO}$ impact the comparison of scaling factors between outflows and host galaxies propertie \citep[e.g.,][]{fio17,flu19} by a factor of $\sim$5.

\subsection{Mass outflow rate}

To estimate the mass outflow rate, along with the observational quantities (outflow mass $M_{out}$, size $\rm R_{out}$ and velocity $\rm v_{out}$), we need to assume a certain geometry. Following \citet{fio17} and \citet{cicone14}, for a spherical or multi-conical geometry, in which the outflowing clouds are uniformly distributed along the flow, the mass outflow rate $\dot{M}_{out}$ can be calculated as:
\begin{equation}
\dot{M}_{out} = 3 v_{out} \left(M_{out}/R_{out} \right)
\end{equation}
If instead we assume a time-averaged thin expelled shell geometry \citep{rupke05b}, also adopted in the study of molecular outflows in the local Universe \citep{vei17,flu19}, we have
\begin{equation}
\label{shell}
\dot{M}_{out} =  v_{out} \left(M_{out}/R_{out} \right)
\end{equation}
which corresponds to the outflow mass averaged over the flow timescale, $t_{flow}=\frac{R_{out}}{v_{out}}$. The difference in the mass loading factor between the two proposed scenarios for the outflow geometry is a factor 3 times larger in the multi-conical/spherical description.

In the following estimates, we use Eq.~\ref{shell} to derive more conservative outflow energetics, since the observations cannot constrain the geometry, and thus favour one scenario over the other.

As discussed in Section~\ref{kine}, the outflow is found in a region of $R_{out}$=0\farcs28 ($\sim$23\,pc) and here we will consider the maximum projected velocity of the wings, $v_{out}$=300\,km/s. If the outflow direction is between the observer line of sight and the galaxy plane, even assuming the maxima projected velocities, the de-projected velocities would encompass the adopted value, therefore, $v_{out}$=300\,km/s is a conservative value. The flow timescale is then $t_{flow}\sim10^4$\,yr, which is comparable to the timescales of the BH growth bursts episodes of nuclear activity, with a duration of $10^{4-5}$\,yr \citep{wada04}. 

For an outflow mass of $M_{out}$=(1.9-2.2)$\times$10$^6$\,M$_\odot$ (assuming the standard $\rm \alpha_{CO,MW}$), we find a mass outflow rate of $\dot{M}_{out}$=(25-29)\,$\rm M_\odot yr^{-1}$. If instead, we use the mass derived assuming the typical values for galaxy centers, $\rm \alpha_{CO}=\frac{1}{4}\alpha_{CO,MW}$, we find that the mass load rate is $\dot{M}_{out}\sim7\,M_\odot yr^{-1}$.

\subsection{The nuclear molecular outflow in dense gas tracers}

\begin{figure}
 \resizebox{\hsize}{!}{ \includegraphics{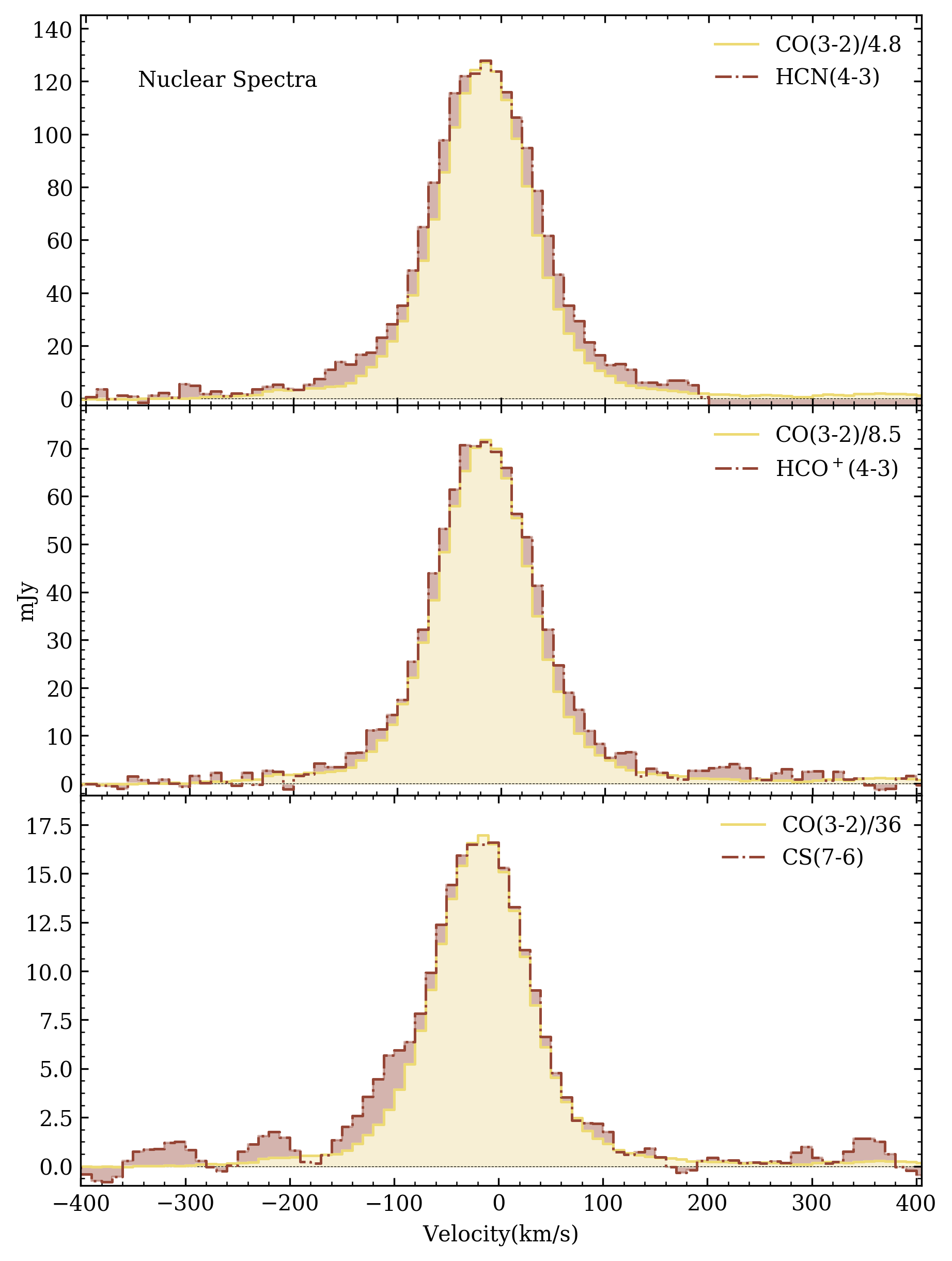}}
     \caption{The nuclear spectra extracted in a $r\sim$0.28\arcsec aperture around the AGN position for HCN(4-3) (\textit{left}), HCO$\rm^+$(4-3) (\textit{middle}) and CS(7-6) (\textit{left}). The dense gas spectra were multiplied by a factor to scale with the CO(3-2) emission in order to compare their wings profiles.} 
     \label{fig:dense_nuc}
\end{figure}

We also detected the presence of broad wings in the nuclear spectrum of dense gas tracers $\rm HCN(4-3)$, $\rm HCO^+(4-3)$ and $\rm CS(7-6)$, as indicated in Figure~\ref{fig:dense_nuc}. We also show the CO nuclear spectrum for comparison, and the high-velocity components cover the same velocity width of the CO wings ($\pm$300\,km/s). We find that the ratio between the peak fluxes for the main \textit{core} in the dense gas and CO are $\sim$5, $\sim$8.5, $\sim$36, for $\rm HCN(4-3)$, $\rm HCO^+(4-3)$ and $\rm CS(7-6)$, respectively. We also can notice in Fig.~\ref{fig:dense_nuc}, there is some indication that, at least for HCN(4-3) and CS(7-6), the line ratios of the blue wings tend to be higher than the core.

\begin{figure}
 \resizebox{\hsize}{!}{ \includegraphics{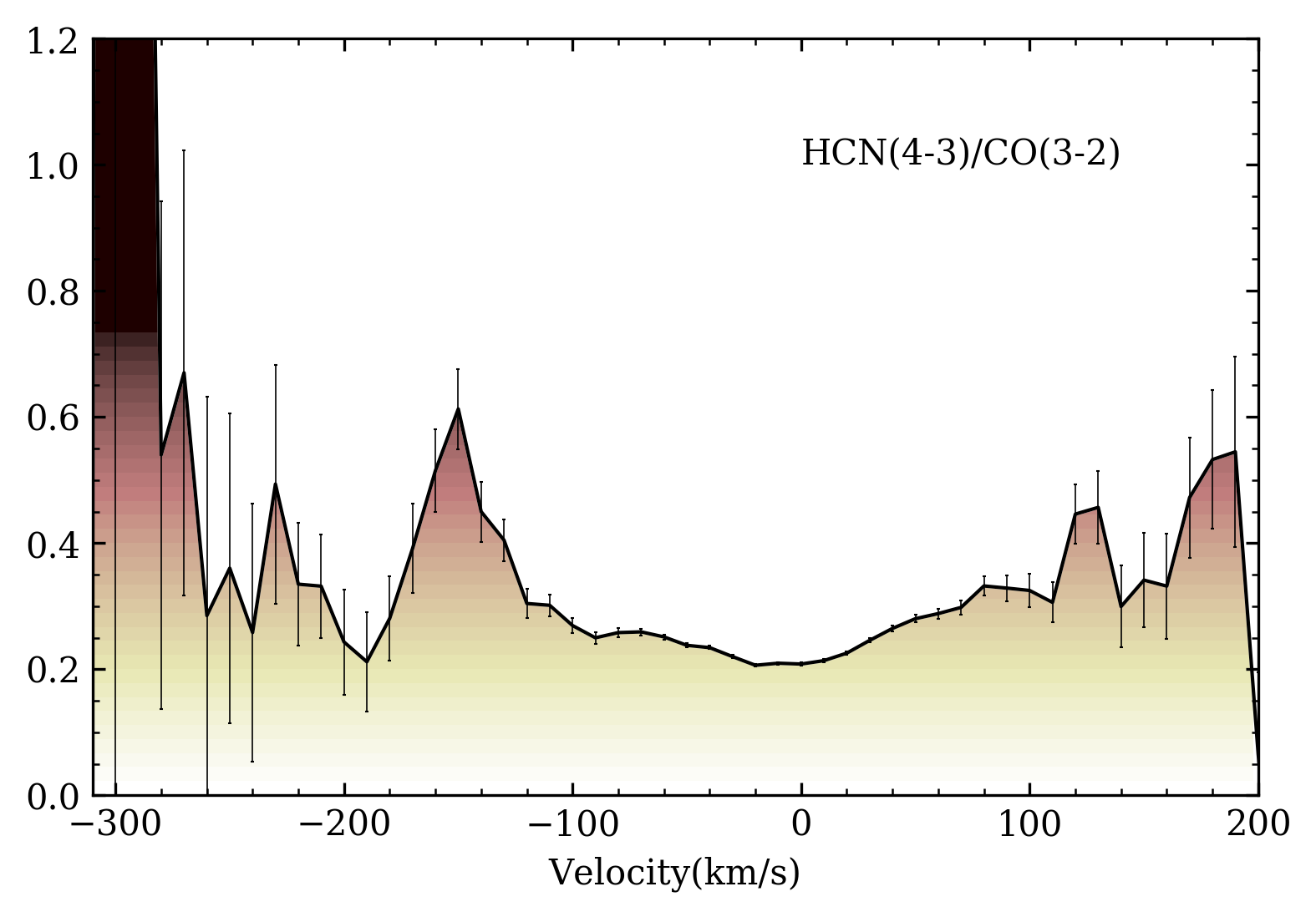}}
     \caption{The HCN(4-3)/CO(3-2) ratio in the nuclear spectra. The ratio is shown up to velocities +200\,km/s due to the tuning of Cycle 3 described in Section~\ref{obs}. We can see that in the core, the ratio is $\sim$0.2 and increases towards the wings up to values $\sim$0.6, indicating that the outflow is entrained mostly in a dense gas ($\rm n\gtrsim10^4\,cm^{-3}$), as discussed in \citet{aalto12}.} 
     \label{fig:ratio}
\end{figure}

In order to quantify the line ratio in the core and wings, we show the $\rm HCN(4-3)/CO(3-2)$ along the nuclear spectra in Figure~\ref{fig:ratio}. The ratio is shown up to velocities +200\,km/s due to the tuning of Cycle 3 described in Section~\ref{obs}. The core component, define by the disk rotation with velocities up to $\pm$100\,km/s, has a ratio of $\sim$0.2 and the ratio increases for the high-velocities towards the wings, up to values of $\sim$0.6, suggesting and enhancement of HCN in the outflow.  As discussed in Section~\ref{dense}, the nuclear region of NGC\,613 presents excitation conditions typical of XDRs in the vicinity of AGN, when analysing the dense gas ratios in the sub-millimeter-HCN diagram  \citep{izumi16} in Figure~\ref{fig:submm}. Yet, we find evidence that the HCN in the outflow can be a factor 3 times higher than the values found in the nuclear CND. 

A similar trend was also reported in the molecular outflow of the QSO galaxy Mrk\,231. The detection of the outflow in HCN(1-0) by \citet{aalto12} covers the same velocity range ($\pm750\,km/s$) of the CO(1-0) outflow \citep{fer10}, and they found a high ratio of the HCN/CO$\sim$0.3-1 in the outflow, higher than in the line core. The HCN is enhanced in the line wings by factors of 2-5, and they suggest that the outflow is mostly entrained in dense gas $n\gtrsim10^4 cm^{-3}$, which is consistent with the molecular gas being compressed and fragmented by shocks \citep{aalto12}. High resolution observations of HCN and HCO$\rm ^+$ in the higher J=3$\rightarrow$2 transition exhibit prominent, spatially extended line wings for HCN(3-2) in Mrk\,231 \citep{aalto15}. In Mrk 231 there were no line wings detected in HCO$\rm ^+$(3-2), while in NGC\,613, there is some indication of high-velocity gas. \citet{aalto15} claimed that the elevated HCN abundance in the outflow is possibly caused by high temperatures in the X-ray irradiated gas regions surrounding AGN \citep{harada13}.

Another possibility to explain the HCN enhancement in the outflow of NGC\,613, is that the HCN emission stems from shocks potentially originated from the interaction of the ouflowing gas with the radio jet. The fact that the molecular outflow is spatially aligned with the central blob of the radio jet detected by \citet{hum92} (see Fig.~\ref{fig:wings}), region where there is evidence of shock excitation as discussed in Section~\ref{nir}, corroborates this scenario. A further detailed analysis of the line ratios for all the NUGA sample will be discussed in a future paper.

\subsection{The driving mechanism: AGN or star formation?}

The origin of the outflow might  be related to star formation, which is concentrated in the nuclear ring region. The star formation rate (SFR) can be estimated from the IR luminosity and the calibration from \citet{kenni98}. From the IRAS fluxes, the IR luminosity is $L_{IR}$=3$\times$10$^9$\,L$_\odot$ \citep[Table \ref{tab:prop},][]{sturm02}, and the total SFR equals to 5.3\,M$_\odot$/yr.  Based on the H$\alpha$ luminosity associated with star formation in the central 3$\times$3\,kpc measured by \citet{davies17}, $\rm L(H\alpha)_{SF}=5.28\times10^{41}$\,erg/s, we can also deduce from Kennicutt's calibration a SFR=4.1\,M$_\odot$/yr, which is consistent with the estimate from IR luminosity. If we isolate the contribution only coming from circumnuclear star-forming ring, the SFR estimated is SFR=2.2\,M$\rm _\odot$/yr in the ring of r$\lesssim$400\,pc \citep{mazzu08}. Evidence of  young star-forming regions in the ``hotspots'' along the ring has been recently found by \citet{fal14}. However, the value reported for NGC\,613 in the nuclear region of aperture of $r\sim$40\,pc is very low, SFR$\sim$0.015\,M$\rm _\odot$/yr \citep{fal14}. This low value can be due to extinction factors or a a possible AGN contribution. The same nuclear region has a large reservoir of warm molecular gas (e.g., see Fig.\ref{fig:sinf}), also found in other Seyfert galaxies \citep{hicks13}. \citet{fal14} suggest a cyclical episode of starburst about $\sim$10\,Myr ago, followed by another episode of nuclear activity. 

We estimate for the nuclear molecular outflow a mass rate of $\rm \dot{M}_{out}\sim27\,M_\odot yr^{-1}$. Although this estimate is uncertain by a factor of a few, given the unknown projection and the assumptions previously discussed in the text, the SFR in the nuclear region is about 3 orders of magnitude lower than $\rm \dot{M}_{out}$.  In general, galactic winds driven by starbursts correspond to mass outflows rates of the same order as the SFR \citep[e.g.,][]{vei05}. Given the discrepancy between the SFR in the nuclear region of and the mass load rate of the outflow, we conclude that star formation alone is not able to drive the nuclear molecular outflow in NGC\,613.

It has been already established that the mass outflow rate increases with the AGN luminosity, supporting the idea of a luminous AGN pushing away the surrounding gas through a fast wind. Previous observational works \citep{cicone14,carni15,fio17} have shown that the molecular outflow properties are correlated with the AGN luminosity, where the outflow kinetic power corresponds to about 5\%$\rm L_{AGN}$ and the momentum rate is $\sim$20$\rm L_{AGN}$/c, in agreement with theoretical models of AGN feedback \citep{fau12, zub12}. For a sample of molecular and ionised outflows, \citet{carni15} found that the ionised gas only traces a small fraction of the total gas mass, suggesting that the molecular phase dominates the outflow mass. This trend is also found by \citet{fio17}, but the ratio between molecular to ionised mass outflow rates is reduced at the highest AGN bolometric luminosities. However, the authors have analysed different samples of galaxies, and this conclusion could be affected by selection bias.

From \textit{XMM-Newton} observations of NGC\,613, \citet{casta13} reported a X-ray luminosity of log$L_{X}(2-10keV)=$41.3\,erg/s.  Applying a bolometric correction from \citet{marco04} gives an AGN bolometric luminosity of $\rm L_{AGN,X}=1.7\times10^{42}$\,erg/s. The bolometric luminosity derived by \citet{davies17} using the [O\textsc{iii}] emission associated only with the AGN contribution, traced as an extended ionization cone aligned with the radio jet, is $\rm L_{AGN,[OIII]}=4\times10^{42}$\,erg/s. If we include the shock and star forming contributions of the total \o3 emission, it gives $\rm L_{bol,[OIII]}=3.75\times10^{43}$\,erg/s. The shock contribution most likely arises from the radio jet launched by the AGN, but here we cannot disentangle the contribution from star formation;
the latter probably overestimates the bolometric AGN luminosity, while the former probably sets a lower limit. 

In a recent study, \citet{flu19} have identified 45 molecular outflows in the local Universe using previous results from the literature and new detections from ALMA archive. They propose an even tighter empirical relation between the mass outflow rate and the SFR, stellar mass, $M_*$, and the bolometric $L_{AGN}$:
\begin{eqnarray}\label{flu}
\log(\dot{M}_{out})=1.14 \log \left( 0.52 \frac{SFR}{\mathrm{M_\odot yr^{-1}}} +0.51 \frac{L_{AGN}}{\mathrm{10^{43}erg.s^{-1}}} \right)  \\
 -0.41 \log \left( \frac{M_{*}}{\mathrm{10^{11}M_\odot}} \right) \nonumber
\end{eqnarray} 
where the SFR is calculated from the total IR luminosity and $\dot{M}_{out}$ is in $\mathrm{M_\odot yr^{-1}}$. We adopt the bolometric AGN luminosity derived from the \o3 emission line, $L_{AGN}=4\times10^{42}$\,erg/s \citep{davies17}, the total SFR inferred from the IRAS fluxes, SFR=5.3$\rm M_\odot yr^{-1}$ and the stellar mass of $M_*=4.5\times10^{10}M_\odot$ calculated in Paper I, derived from the S$\rm ^4$G3.6\,$\mu$m IR image and the GALFIT decomposition \citep{salo15}. Hence, according to Eq.~\ref{flu}, we should expect an outflow mass rate of $4.8\,M_\odot yr^{-1}$ in NGC\,613, which corresponds to a factor of 6 times lower than our estimate. 

The kinetic power of the nuclear outflow can be estimated as $P_{K,out}=0.5 v^2\dot{M}_{out}$. For the $\alpha_{CO,MW}$ assumption, $\rm \dot{M}_{out}\sim27\,M_\odot yr^{-1}$, and we find $P_{K,out}=8\times10^{41}$\,erg/s, which corresponds to 20\%$L_{AGN}$. This value exceeds the predictions from AGN feedback models and cosmological simulations that require that a fraction of the radiated luminosity should be coupled to the surrounding gas $\sim5\%L_{AGN}$ \citep{dimat05,zub12}. However, these predictions are based on powerful AGN, accreting close to the Eddington limit and we hypothesise that the coupling efficiency between AGN-driven outflows and $L_{AGN}$ should be outweighed in LLAGN. The Eddington luminosity derived from the BH mass ($M_{BH}$=3.7$\times10^{7}M_\odot$, Paper I) is $L_{Edd}=4.6\times10^{45}$erg/s, leading to a low accretion rate $L_{AGN}/L_{Edd}\lesssim1\times10^{-3}$. If we assume the lower $\alpha_{CO}=1/4\alpha_{CO,MW}$, the kinetic power of the outflow is $P_{K,out}\sim2\times10^{41}$\,erg/s, and the coupling is 5\%$L_{AGN}$. In both assumptions, the results indicate that the AGN can power the nuclear outflow in NGC\,613, but the former requires a higher coupling efficiency.

The momentum flux of the outflow can be computed from $P_{out}=\dot{M}_{out} v\sim5\times10^{34}$\,dynes. Compared to the momentum provided by the AGN photons,  $L_{ AGN}/c$=1.3$\times10^{32}$\,dynes, it is higher by a factor of $\dot{M}_{out}v\sim400L_{AGN}/c$. In case of energy-conserved winds, AGN feedback models predict a momentum boost by factors up to 50 \citep[e.g.][]{fau12}. Even assuming the lower $\alpha_{CO}$, the value would exceed the predictions ($\sim100L_{AGN}/c$). One possibility to explain the high apparent energetics of the outflow in NGC\,613 is that the AGN activity was stronger in the past. As discussed by \citet{flu19}, they found that 10\% present outflows that exceed the theoretical predictions. They suggest these galaxies could have \textit{fossil outflows}, resulting from a strong past AGN that now has already faded. In the case of the LLAGN NGC\,1377, a collimated molecular outflow is detected at 150\,pc scales \citep{aalto16}, possibly entrained by a faint radio jet, and the authors suggested that the nuclear activity of NGC\,1377 may also be fading. 
 
Alternatively, it is possible that the outflow is driven through the AGN radio jets. The radio power of the jet can be estimated from the 1.4\,GHz luminosity, $P_{1.4}$, using the relation of \citet{birzan08}. If we compute the $P_{1.4}$ luminosity using the NRAO VLA Sky Survey (NVSS) flux measurement of 179.6\,mJy at $\nu$=1.4\,GHz\citep{condon98}, we find a radio power $P_{jet}=1.2\times10^{43}$\,erg/s. Since this value is very elevated, it might include emission from the circumnuclear star-forming ring. To avoid this contribution, we used only the flux of the linear component of the radio jet at 4.86\,GHz by \citet{hum92}. At 4.86\,GHz, the flux associated to the jet is $S_\nu$=7.2\,mJy, and using a spectral index of $\alpha$=-0.8 to derive the flux at 1.4\,GHz, we find $P_{jet}=5.5\times10^{42}$\,erg/s. Hydrodynamical simulations of the interaction of the jet with a clumpy interstellar medium has shown that the jet is able to drive a flow efficiently as soon as the Eddington ratio of the jet P$_{jet}$/L$_{Edd}$ is larger than 10$^{-4}$ \citep{wagner12}. This condition is required for the jet-driven outflow velocity to exceed the velocity dispersion of the $M-\sigma$ relation but it also depends on the cloud sizes in the ISM. In NGC\,613, this ratio is about 1.2$\times$10$^{-3}$, and the jet power is about one or two orders of magnitudes higher than the kinetic luminosity of the outflow. We conclude that the jet is able to drive the molecular outflow, even with low coupling.

The molecular outflow in NGC\,613 is an intriguing case where a very powerful molecular outflow is detected in a LLAGN. The SFR is very weak in the nuclear region, and therefore, not able to drive the flow. Radio jets are found to play a role in the driven mechanism to accelerate the molecular gas in other LLAGN. This is the case for the LINERs NGC\,1266 and NGC\,6764, as suggested by \citet{ala11} and \citet{leon07}, respectively, and the Seyfert 2 NGC\,1433 \citep{combes1433}. The properties of the flow require the contribution of the AGN through the entrainment of its radio jets.

%In particular, the spectrum of NGC 0613 reveals a very broad emission feature (FWHM of 87 km s), and such broad lines have been typically associated with radio jets rather than molecular disks (e.g., Peck et al. 2003; Claussen et al. 1998).  kondrakto+2006

\section{Torques and AGN fueling}\label{torque}

In Paper I, we reported the ALMA observations of the CO(3-2) line for all the galaxies in the NUGA sample. In three cases, NGC\,613, NGC\,1566 and NGC\,1808, the CO emission has revealed a nuclear trailing spiral, as the large-scale one. Inside the nuclear ring at the ILR of the bar, usually a leading spiral is expected, developing transiently, and generating positive torques, which drive the inner gas onto the ring. However, when the gravitational impact of the black hole is significant, the spiral can then be trailing, and the torques negative, to fuel the nucleus \citep{buta96,fukuda98}.

We find filamentary structures in NGC\,613, pointing to a radial gas transport from the circumnuclear star forming ring into the core region dominated by the AGN, as displayed in Figure~\ref{fig:mom_613}. The ring morphology appears disturbed by a radial outflow of material from the AGN, which is confirmed by the existence of a weak jet in archival radio maps. However, the radio jet does not seem to have any significant effect on the morphology of the large reservoir of molecular gas that has accumulated inside the central 100 pc. 

If gas is inflowing from the bar dust lanes into the ring, as expected from gravity torques \citep{santi05}, there is also an inflow in the CND, due to the nuclear trailing spiral, as already observed for NGC\,1566 \citep{combes1566}. Inside the nuclear spiral structure, there is a very dense and compact (radius $\sim$ 14~pc) rotating component, which might be interpreted as the molecular torus (Paper I). 

In order to explore the efficiency of feeding in NGC\,613, we have estimated the gravitational torques exerted by the stellar potential on the molecular gas, following the methodology described by \citet{santi05}. The gravitational potential is computed in the plane of the galaxy using a red image (F814W) from HST, since 2MASS images have insufficient angular resolution. We have not separated the bulge from the disk contribution since NGC\,613 is a late type (Sbc) galaxy. This means that the bulge was effectively assumed to be flattened. Dark matter can be safely neglected inside the central  kpc. The image has been rotated and deprojected according to PA=120$^\circ$ and i=41$^\circ$, and then Fourier transformed to compute the gravitational potential and forces. A stellar exponential disk thickness of $\sim$1/12th of the radial scale-length of the galaxy (h$_{\rm r}$=3.8kpc) has been assumed, giving h$_{\rm z}$=317pc. This is the average scale ratio for galaxies of this type \citep[e.g.,][]{barna92,biz09}. The potential has been obtained assuming a constant mass-to-light ratio of M/L=0.5 in the I-band over the considered portion of the image of 1\,kpc in size. This value is realistic in view of what is found statistically for spiral galaxies \citep{bell01}. The pixel size of the map is 0.06\arcsec=5\,pc, the average value between the  ALMA and HST resolutions. The stellar M/L value was fit to reproduce the observed CO rotation curve. 

\begin{figure}
\resizebox{\hsize}{!}{\includegraphics{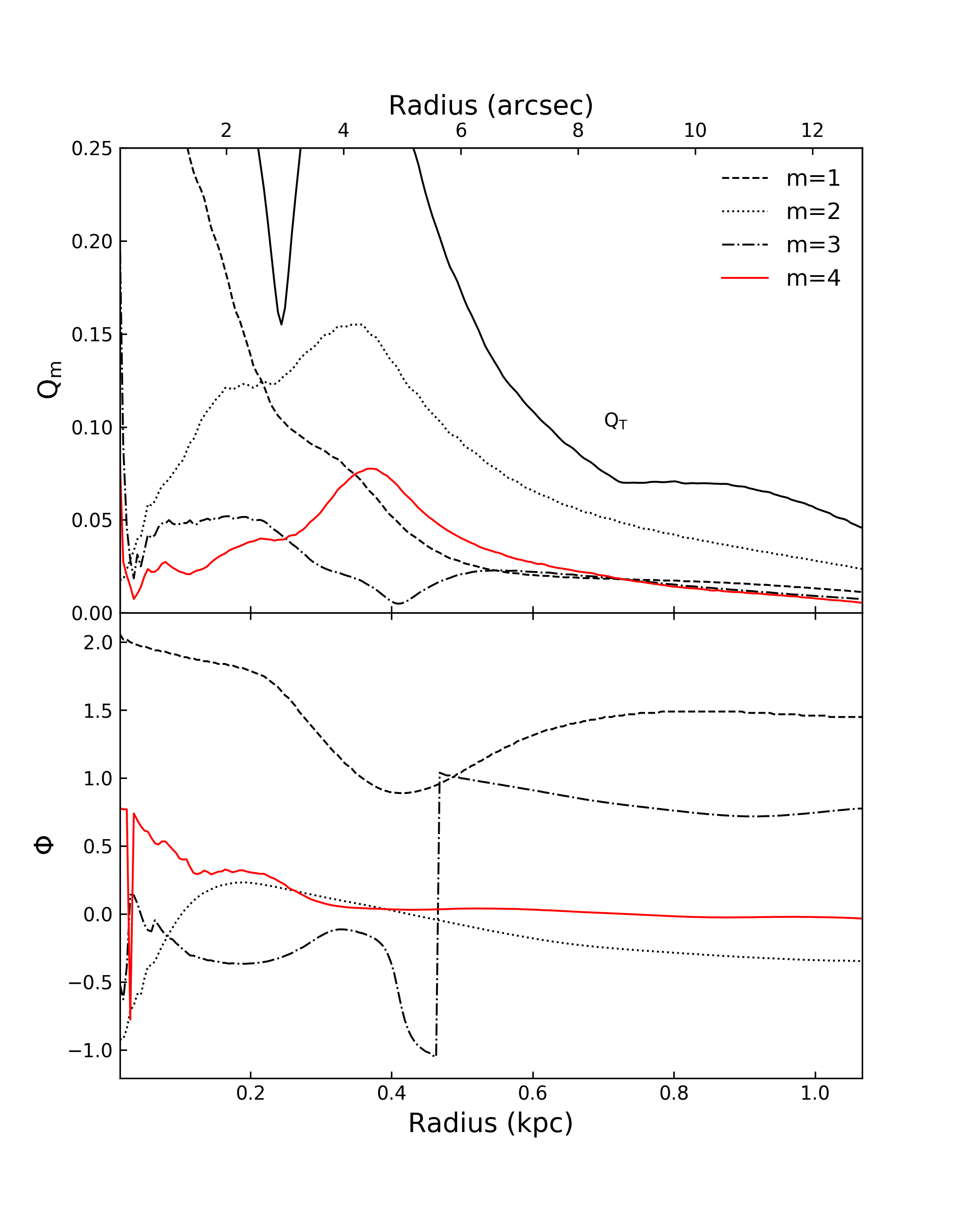}}
\caption{{\it Top} Strengths (Q$_m$  and total Q$_T$) of  the $m=1$ to $m=4$ Fourier components of the stellar potential within the central kpc. The $m=1$ term is dominant up to a radius of 200\,pc, and after the m=2 component dominates and has a constant phase, corresponding to the bar. {\it Bottom} Corresponding phases in radians of the Fourier components, taken from the major axis, in the deprojected image.}
\label{fig:qpot}
\end{figure}

The potential $\Phi(R,\theta)$ can be decomposed into its different Fourier components:
\begin{equation}
 \Phi(R,\theta) = \Phi_0(R) + \sum_m \Phi_m(R) \cos (m \theta - \phi_m(R))
\end{equation}
\noindent
where $\Phi_m(R)$ and $\phi_m(R)$ are the amplitude and the phase of the $m-$mode, respectively. The strength of the $m$-Fourier component, $Q_m(R)$ is defined as $Q_m(R)=m \Phi_m / R | F_0(R) |$, i.e. by the ratio between tangential and radial forces \citep{combes81}. The strength of the total non-axisymmetric perturbation $Q_T(R)$ is defined  similarly with the maximum amplitude of the tangential force $F_T^{max}(R)$. Their radial distributions and the radial phase variations are displayed in Fig.~\ref{fig:qpot}.

\begin{figure}
\resizebox{\hsize}{!}{\includegraphics{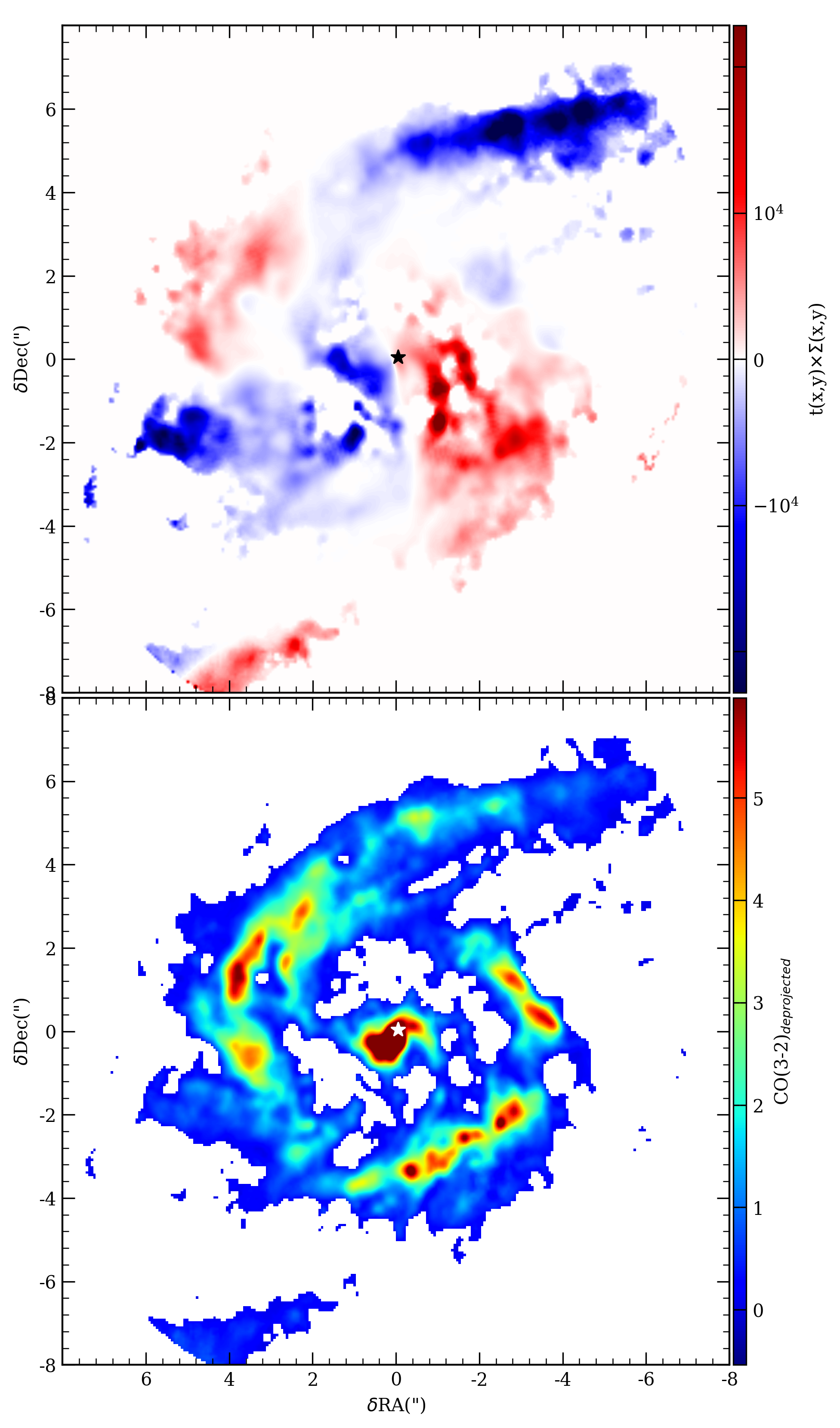}}
\caption{{\it Top:} Map of the gravitational torque,  (t(x,y)~$\times$~$\Sigma$(x,y), as defined in the text) in the center of NGC\,613.The torques change sign as expected in a four-quadrant pattern (or butterfly diagram). The orientation of the quadrants follows the nuclear bar's orientation. In this deprojected picture, the major axis of the galaxy is oriented parallel to the horizontal axis. {\it Bottom:} The deprojected image of the CO(3-2) emission, at the same scale, and with the same orientation, for comparison. The color scales are linear, in arbitrary units.}
\label{fig:torq1}
\end{figure}

The derivatives of the potential yield the forces per unit mass ($F_x$ and $F_y$) at each pixel, and the torques per unit mass $t(x,y)$ are then computed by $t(x,y) = x~F_y -y~F_x$. The sign of the torque is determined relative to the sense of rotation in the plane of the galaxy. The product of the torque and the gas density $\Sigma$ at each pixel allows one then to derive the net effect on the gas, at each radius.  This quantity $t(x,y)\times\Sigma(x,y)$, is shown in Fig.~\ref{fig:torq1}, together with the deprojected CO map.

\begin{figure}
\resizebox{\hsize}{!}{\includegraphics{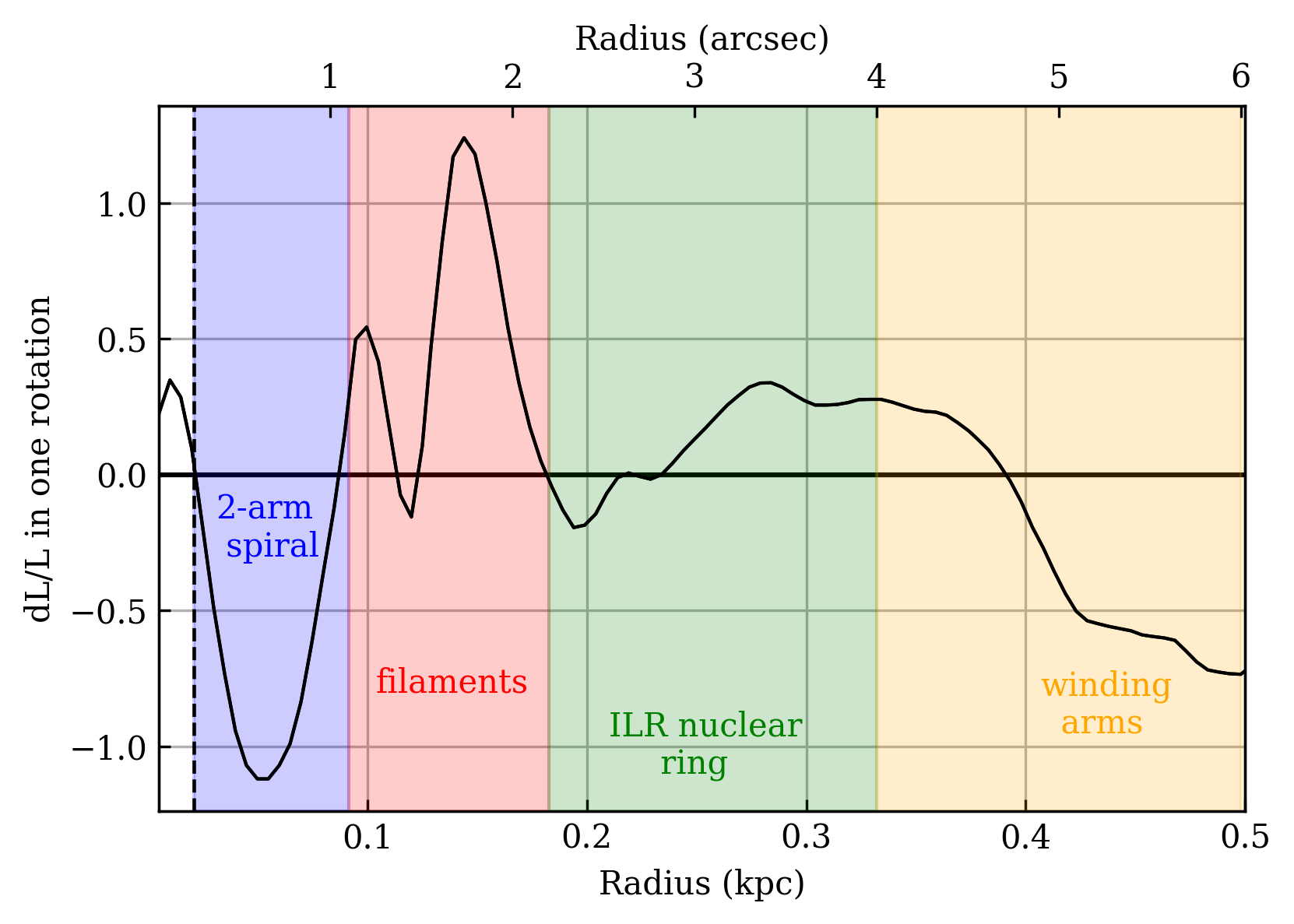}}
\caption{The radial distribution of the torque, quantified by the fraction of the angular momentum transferred from the gas in one rotation--$dL/L$, estimated from the CO(3-2) deprojected map. The vertical dashed line at 25\,pc radius delimitates the extent of the central gas outflow, and the computation has no meaning here. The torque is negative inside the $\lesssim$100\,pc nuclear spiral and in the winding arms and positive in the filaments.}
\label{fig:gastor}
\end{figure}

The torque weighted by the gas density $\Sigma(x,y)$ is then averaged over azimuth, i.e.
\begin{equation}
  t(R) = \frac{\int_\theta \Sigma(x,y)\times(x~F_y -y~F_x)}{\int_\theta \Sigma(x,y)}
\end{equation}
 \noindent
The quantity $t(R)$ represents the time derivative of the specific angular momentum $L$ of the gas averaged azimuthally \citep{santi05}. Normalising at each radius by the angular momentum and rotation period ($\rm T_{rot}$) allows us to estimate the efficiency of the gas flow, as shown in Fig.~\ref{fig:gastor}. The torques are negative in the winding arms at r$\sim$500\,pc (corresponding to the dust lanes) and are indeed contributing to drive the outer gas into the star-forming ring at $\sim$350\,pc. The observations of nuclear rings are more common among barred galaxies \citep{kor04,pee06,jogee06}, and can be explain by the  gas slowing down as it crosses the ILR, consequently weakening the gravitational torques, and the gas piles up in rings \citep{combes85, byrd94}. We can see in Fig.~\ref{fig:gastor} that the efficiency certainly  drops in the inner ILR region. The filaments within the nuclear region, between 100 and 200\,pc, show that the gas gains angular momentum in one rotation, which is $\rm T_{rot}\sim$12\,Myr.  

The nuclear bar strength is moderate, however, the BH has a strong influence on the nuclear molecular gas, as shown in Fig.~\ref{fig:gastor}, the fueling efficiency is high in the nuclear spiral. Between 25 and 100\,pc, the gas loses its angular momentum in one rotation, which is $T_{rot}(100pc)\sim$9.5\,Myr. Inside 25\,pc, the torques are positive and correspond to the region where we detect the molecular outflow.  Since the gas in this region is not in quasi-stationary orbits, but rather ejected from the galaxy plane, the computation cannot be interpreted in terms of the average torque here.

As shown in Fig.~\ref{fig:torq1}, the nuclear spiral structure inside the ILR ring 
of the bar is of a trailing nature and is located inside the negative torque quadrants. This  
might appear surprising, since in many cases the spiral structure is predicted to be leading
in this region, and the torque positive, maintaining the gas in the ILR ring \citep{buta96}. However, in  the presence of a sufficiently massive black hole, this behaviour can be reversed: the spiral becomes trailing, and the torque negative. Indeed, the indicator of the precession rate of elliptical orbits in the frame of the epicyclic approximation, $\Omega-\kappa/2$, is significantly modified by a central massive body. Instead of decreasing regularly towards zero at the center, the $\Omega-\kappa/2$  curve increases steeply as r$^{-3/2}$. The gas undergoes collisions, loses energy and spirals progressively towards the center. When the precession rate decreases, the series of elliptical orbits precess more and more slowly, and then lag at smaller radii,  forming a leading structure. When the precession rate increases, they form a trailing structure \citep{combes1566}.

One way to detect spiral structure in the nuclear gas is to amplify the dust extinction features in the HST images. However, we would like to stress that observations of spiral dust lanes interior to the star-formation rings are not very common \citep{martini03,pee06}.
Notwithstanding, a clear case is seen in NGC\,2207, where dusty spirals extend from $\sim$50 to 300\,pc, suggesting that the gas continues to sink inside the ILR and might be promoting gas accretion into the nucleus \citep{elme98}. Multiple spiral arms were also detected inside the star-forming ring ($r\sim$700\,pc) in the LINER/Sy\,1 galaxy NGC\,1097 by \citet{prieto05}. They compared their results with hydrodynamic models of \citet{macie04}, concluding that inflows were occurring along the nuclear spirals. Although nuclear star forming rings are common, their interior structure is difficult to distinguish in dust maps, making the molecular gas emission a better tracer. The NUGA maps show clear evidence of trailing spirals in three objects of the sample, highlighting the importance of the high resolution ALMA observations.
 
%Another question is whether the orbits change orientation at the ILR, as expected if the x$_2$ orbits develop between two inner resonances. This would destroy the nuclear bar and be observable on the infrared image. The lower panel of Fig.~\ref{fig:qpot} shows the phases of the various Fourier components. The bar ($m=2$) has a constant phase over its length, and their orientation change does not appear to be significant.

NGC\,613 is therefore an example of  a trailing spiral inside the nuclear ring of a bar i.e., the case described in \citet{buta96}. This means that the mass of the black hole should be sufficiently high to have an influence on the gas dynamics on a 100\,pc scale. In summary, the gravity torques are negative in the winding spirals, and the gas accumulates in the star forming ring at the inner ILR of the nuclear bar. The filamentary structure gains angular momentum and then the nuclear spiral at $\lesssim$100\,pc present a very high efficiency in fuelling the central BH. 

%\begin{figure}
%  \resizebox{\hsize}{!}{\includegraphics{omega.png}}
%  \caption{Schematic representation of the rotation curve (black)
%epicyclic frequency $\kappa$ (red), and corresponding $\Omega-\kappa/2$ curve (blue)
%within the central kpc of NGC\,613. The contribution of a super-massive black
%hole in the nucleus with M$_{\rm BH}$ = 8.3 10$^6$ \msol\ has been included.}
%\label{fig:vbh}
%\end{figure}

\section{Conclusions}\label{con}

We have presented the combined ALMA cycle 3 and 4 observations for the Seyfert/nuclear starburst galaxy NGC\,613. The combined observations in CO(3-2) reach a spatial resolution of 0.2\arcsec$\sim$17\,pc. We study the morphology and the kinematics of the molecular gas in the central 1\,kpc and our main findings are summarized below:

$\bullet$ The 350\,GHz continuum map shows a compact, barely resolved, emission peak at the position of the AGN, surrounded by a patchy ring, matching the star-forming ring seen on optical images.

$\bullet$ The morphology of CO(3-2) line emission reveals several components: a 2-arm nuclear spiral at $\rm r\lesssim$100\,pc trailing the gas toward the center, a circumnuclear ring $\sim$350\,pc, that correspond to the star-forming ring. Also, we find evidence of a filamentary structure connecting the ring and the nuclear spiral. The ring reveals two breaks into two winding spiral arms, at NW and SE, corresponding to the dust lanes in the HST images.

$\bullet$ The kinematics of the CO emission show a rather regular rotational velocity field in the inner kpc disk. We applied a tilted-ring model to fit the velocity map and the residuals show indeed the bulk of the molecular gas in circular motion except in the west part of the ring, which perturbations are due to the contact point between the ring and the SE winding arm.

$\bullet$ There is a remarkable coincidence between the molecular gas and the warm H$_2$ and ionised gas, traced by the [Fe\textsc{ii}] and Br$\gamma$ emission lines, in the star-forming ring. Line diagnostics in the NIR indicate that the clumps in the ring are in agreement with young star-forming excitation. On the other hand, the nucleus of NGC\,613 presents an excitation mechanism typical of Seyfert or LINER. 

$\bullet$ We measured the line intensity ratios R$_{\rm HCN/HCO^{+}}$ and R$_{\rm HCN/CS}$ in the nuclear region and in a clump along the ring. We find that the ratio for the nuclear region points to the AGN-dominated part of the HCN-submillimeter diagram, while the ratio for the clump is located in the starburst-dominated part. These results indicate that the nuclear region of NGC\,613 presents line ratios in agreement with excitation conditions typical of XDRs in the vicinity of AGN.

$\bullet$ In the PV diagrams, we find skewed kinematics in the nuclear region of r$\sim$25\,pc. This feature is seen as broad wings ($v\pm$300\,km/s) in the CO nuclear spectrum, and the wings are also present in the dense gas tracers. We identify this feature as a molecular outflow emanating from the nucleus. The molecular outflow is co-spatial with the central blob detected in the radio jet.

$\bullet$ We derive the molecular mass associated to the outflow as $M_{out}$=2$\times$10$^6$\,M$_\odot$. The mass outflow rate is $\dot{M}_{out}=$27\,$\rm M_\odot yr^{-1}$. If instead, we use the mass derived assuming the typical values for galaxy centers, $\rm \alpha_{CO}=\frac{1}{4}\alpha_{CO,MW}$, we find that the mass load rate is $\dot{M}_{out}\sim7\,M_\odot yr^{-1}$.

$\bullet$ We find a HCN enhancement in the outflow, probed by an increasing of the HCN(4-3)/CO(3-2) ratio along the wings in the nuclear spectra. While the core has ratios of about $\sim$0.2, this value increases in the wings up to $\sim$0.6, indicating that the outflow is entrained mostly in a dense gas ($\rm n\gtrsim10^4\,cm^{-3}$). Another possibility is that the HCN emission stems from shocks potentially originated from the interaction of the ouflowing gas with the radio jet.

$\bullet$ The molecular outflow energetics exceed the values predicted by AGN feedback models. The kinetic power of the nuclear outflow corresponds to $P_{K,out}=$20\%$L_{AGN}$ and the momentum rate is $\dot{M}_{out} v\sim400L_{AGN/c}$. We speculate that, given its current weak nuclear activity, NGC\,613 might be a case of \textit{fossil outflows}, resulted from a strong past AGN that now has already faded. 

$\bullet$ The outflow can be entrained by its radio jet. We find that P$_{jet}$/L$_{Edd}\sim$10$^{-3}$ and the jet power is about one or two orders of magnitudes higher than the kinetic luminosity of the outflow. In these conditions, the  jet is able to drive the molecular outflow.

$\bullet$ The trailing spiral observed in CO emission is inside the ILR ring of the bar. We have computed the gravitational potential from the stars within the central kpc, from the I-band HST image. Weighting the torques on each pixel by the gas surface density observed in the CO(3-2) line has allowed us to estimate the sense of the angular momentum exchange and its efficiency. The gravity torques are negative from 25 to 100\,pc. Between 50 pc and 100\,pc, the gas loses its angular momentum in a rotation period, providing evidence of fueling the AGN.

The molecular outflow in NGC\,613 is an intriguing case where a very powerful molecular outflow is detected in a LLAGN. The SFR is very weak in the nuclear region, and therefore, not able to drive the flow. The properties of the flow require the contribution of the AGN through the entrainment of its radio jets. On the other hand, there is a clear trailing spiral observed in molecular gas inside the ILR ring of a bar, indicating that the super-massive black hole is influencing the gas dynamics. Instead of maintaining the ILR ring density, the torques are then driving gas towards the nucleus, a first step towards possibly fueling the AGN. NGC\,613 is a remarkable example of the complexity of fuelling and feedback mechanisms in AGN, and reinforce the importance of detailed analysis of nearby galaxies with ALMA capabilities to shed light on the gas flow cycle in AGN. 

\begin{acknowledgements}

%  We warmly thank the referee for constructive comments and suggestions. 
The ALMA staff in Chile and ARC-people at IRAM are gratefully acknowledged for their help in the data reduction. We particularly thank Philippe Salom\'e for useful advice and Jes{\'u}s Falc{\'o}n-Barroso for the SINFONI and VLA data. AA would like to thank the Brazilian scholarship program Science without Borders - CNPq (reference [234043/2014-8]) for financial support. SGB acknowledges support from the Spanish MINECO grant AYA2016-76682-C3-2-P. LH is grateful to INAF PRIN-SKA funding program 1.05.01.88.04. This paper makes use of the following ALMA data: ADS/JAO.ALMA\#2015.0.00404.S, and ADS/JAO.ALMA\#2016.0.00296.S.  ALMA is a partnership of ESO (representing its member states), NSF (USA) and NINS (Japan), together with NRC (Canada) and NSC and ASIAA (Taiwan), in cooperation with the Republic of Chile. The Joint ALMA Observatory is operated by ESO, AUI/NRAO and NAOJ. The National Radio Astronomy Observatory is a facility of the National Science Foundation operated under cooperative agreement by Associated Universities, Inc. We used observations made with the NASA/ESA Hubble Space Telescope, and obtained  from the Hubble Legacy Archive, which is a collaboration between the Space Telescope  Science Institute (STScI/NASA), the Space Telescope European Coordinating Facility (ST-ECF/ESA), and the Canadian Astronomy Data Centre (CADC/NRC/CSA). We made use of the NASA/IPAC Extragalactic Database (NED), and of the HyperLeda database. This research made use of Astropy, a community developed core Python package for Astronomy. This work was supported by the Programme National Cosmology et Galaxies (PNCG) of CNRS/INSU with INP and IN2P3, co-funded by CEA and CNES.
 
\end{acknowledgements}

%-------------------------------------------------------------------
\bibliographystyle{aa} % style aa.bst
\bibliography{test.bib}

\end{document}